\documentclass[%
preprintm, superscriptaddress, 12pt,
 amsmath,amssymb,
 aps,
]{revtex4-2}

\usepackage{scalerel}
\usepackage{siunitx}
\usepackage{dcolumn} 
\usepackage{titlesec}
\titlespacing*{\section}
{0pt}{4.5ex plus 1ex minus .2ex}{1.3ex plus .2ex}
\titleformat{\section}
  {\centering\bfseries\MakeUppercase}{\thesection}{1em}{}

\usepackage{xstring}
\usepackage{tikz}

\usepackage{fancyhdr} 
\usepackage{titlesec}
\titlespacing*{\section}
{0pt}{4.5ex plus 1ex minus .2ex}{1.3ex plus .2ex}
\titleformat{\section}
  {\centering\bfseries\MakeUppercase}{\thesection}{1em}{}

\usepackage{graphicx}
\usepackage{dcolumn}
\usepackage{float}

\usepackage{mathtools}
\usepackage{bm}

\newcommand{\SM}{Supplementary Material}

\usepackage{natbib}
\usepackage[english]{babel}
\usepackage[utf8]{inputenc}
\usepackage{url}
\usepackage[colorlinks = true,
            linkcolor = blue,
            urlcolor  = blue,
            citecolor = blue,
            anchorcolor = blue]{hyperref}

\begin{document}

\title{Transient strain induced electronic structure modulation in a semiconducting polymer imaged by scanning ultrafast electron microscopy}

\author{Taeyong Kim}
\affiliation{Department of Mechanical Engineering, University of California, Santa Barbara, CA 93106, USA}

\author{Saejin Oh}
\affiliation{Department of Chemistry and Biochemistry, University of California, Santa Barbara, CA 93106, USA}

\author{Usama Choudhry}
\affiliation{Department of Mechanical Engineering, University of California, Santa Barbara, CA 93106, USA}

\author{Carl Meinhart}
\affiliation{Department of Mechanical Engineering, University of California, Santa Barbara, CA 93106, USA}

\author{Michael L. Chabinyc}
\email{chabiny@ucsb.edu }
\affiliation{Materials Department, University of California, Santa Barbara, CA 93106, USA}

\author{Bolin Liao}
\email{bliao@ucsb.edu} \affiliation{Department of Mechanical Engineering, University of California, Santa Barbara, CA 93106, USA}

\date{\today}

\begin{abstract}
Understanding the opto-electronic properties of semiconducting polymers under external strain is essential for their applications in flexible opto-electronic, light-emitting and photovoltaic devices. While prior studies have highlighted the impact of static strains applied on a macroscopic length scale, assessing the effect of a local transient deformation before structural relaxation occurs is challenging due to the required high spatio-temporal resolution.
 Here, we employ scanning ultrafast electron microscopy (SUEM) to image the dynamical effect of a photo-induced transient strain in the archetypal semiconducting polymer poly(3-hexylthiophene) (P3HT). We observe that the photo-induced SUEM contrast, corresponding to the local change of secondary electron emission, exhibits a ring-shaped spatial profile with a rise time of $\sim 300$ ps, beyond which the profile persists in the absence of a spatial diffusion. We attribute the observation to the electronic structure modulation of P3HT caused by a photo-induced strain field owing to its relatively low modulus and strong electron-lattice coupling, as supported by a finite-element analysis. Our work provides insights into tailoring opto-electronic properties using transient mechanical deformation in semiconducting polymers, and demonstrates the versatility of SUEM to study photo-physical processes in diverse materials. 
\end{abstract}

\maketitle
\section{Introduction}

Semiconducting polymers are enabling materials for applications such as flexible opto-electronics~\cite{yang2019electronic, someya2017imperceptible}, solar cells~\cite{yan2018non}, and bioelectronics~\cite{inal2018conjugated}. The optical, electrical and mechanical properties of semiconducting polymers can be controlled by judicious design of their conjugated backbone and their sidechains~\cite{MeiandZBao_COM}. The $sp^2$ hybridization along the chain allows the electrons to delocalize, leading to drastically improved opto-electronic properties including high charge carrier mobility exceeding~1~cm$^2$~V$^{-1}$s$^{-1}$ in thin film transistors~\cite{paterson2018recent}. Concurrently, the one-dimensional nature of the polymer chains is expected to give rise to strong coupling between the electronic states and the chain lattice, causing the formation of polaronic states upon the injection of charge carriers into a semiconducting polymer~\cite{ghosh2020excitons}. The strong electron-lattice coupling, in combination with the mechanical properties of polymers~\cite{xie2018connecting}, signals the possibility of tuning the opto-electronic properties of semiconducting polymers via mechanical strain applied to the polymer chains. 

The effects of mechanical deformation on semiconducting polymers has been examined both by theory and experiment~\cite{root2017mechanical}. Computational studies show that structural disorder \cite{poelking2013effect} and strain of polymer chains can modify their electronic structure~\cite{menichetti2017strain}.  The charge carrier mobility of semiconducting polymers has been experimentally shown to be modified by external tensile strain due to the change of interatomic distances, conformational changes, and chain alignment at large deformations ~\cite{qian2016stretchable,root2017mechanical}. Compressive strain can have comparable effects with the optical gap of poly(3-hexylthiophene), P3HT~\cite{hess1993photoexcitations}, and other polymers~\cite{schmidtke2007optical} being strongly lowered by an external hydrostatic pressure.  Despite extensive investigations, however, whether these changes can be controlled for beneficial effects is relatively unexplored compared to other classes of semiconductors. Additionally, in a broader context, a detailed understanding of the coupling effect between mechanical strain and electronic structure is of fundamental scientific interest. Compared to their rigid inorganic counterparts, low-modulus semiconducting polymers provide an interesting platform to study the strain effect on the opto-electronic properties of materials.   

One major limitation of most existing strain studies of semiconducting polymers is that the applied strain is on a macroscopic length scale, usually using direct tensile testing, stretchable substrates, or mechanical indentation measurements~\cite{ChenandZBao_Mat, Vijay_JAP_2011, OhandZBao_Nature_2016, Sahabudeen_Natcomm,GasperiniandZBao_MAMO,Turner_AdvFuncMat, Awartani_AdvEnMat,Moon_nl, Moon_ACSPho}. In many experiments, a step of strain is applied to the sample for a relatively long time period~\cite{xie2018connecting}. While in the strained state, there can be structural relaxation in both ordered and amorphous domains of the polymer, which leads to changes in interchain distances and conformations before measurements of electronic properties. On the other hand, extensive time-resolved spectroscopic studies of exciton and charge carrier dynamics from picosecond to microsecond time scales in semiconducting polymers have been performed, such as transient photomodulation~\cite{ShengVardeny_PRB}, time-resolved photoluminescence~\cite{Piris_JPCC}, transient absorption~\cite{albert2011measurement},~transient microwave conductivity~\cite{Dicker_PRB_2004}, and transient grating~\cite{Wenkai_APL} measurements. In these studies, however, the strain effect is typically a byproduct that is not systematically investigated and the conventional optical spectroscopic techniques lack the spatial resolution to pinpoint the effect of complex spatial strain profiles. In this light, to fully characterize the modulation of opto-electronic properties by strain directly applied to polymer chains, an ideal experimental technique would combine the following features: a high temporal resolution to probe the transient strain effect before significant structural relaxation occurs, a high spatial resolution to map local strain profiles, and a high sensitivity to detect the change of the local electronic structure.

Combining the subpicosecond time resolution of ultrafast lasers and the nanometer spatial resolution of high-energy electron beams, recently developed scanning ultrafast electron microscopy (SUEM) is a suitable technique~\cite{Jerry_PNAS,liao2017scanning,KimSandia,BoseandOmar_nl} to directly image the spatial-temporal effect of photo-induced transient strain fields in conducting polymers. In addition to the high spatial-temporal resolution, SUEM also provides extreme surface sensitivity~\cite{Zani_SUEM}, given the typical secondary electron escape length of a few nanometers. With these advantages, SUEM has been employed to directly visualize surface photocarrier transport across a silicon p-n junction~\cite{Najafi_Science}, a two-dimensional $\mathrm{MoS}_2$ band-bending junction~\cite{wong2021spatiotemporal}, unusual hot photocarrier dynamics in silicon~\cite{Bolin_Natnano,najafi2017super} and in black phosphorus~\cite{Bolin_nl}, photocarrier oscillation under strong photoexcitation~\cite{najafi2021carrier}, and trap-mediated recombination~\cite{BoseandOmar_nl}, among other examples primarily in inorganic substances. Najafi et al. recently applied SUEM to image the photoinduced surface acoustic waves in P3HT on nanosecond and millimeter scales~\cite{Najafi_P3HT}. However, imaging the local impact of photo-induced transient strain on the electronic structure of conducting polymers on smaller time and length scales (picosecond and micrometer) has not been reported before.

 Here, we demonstrate the space-time mapping of the dynamical effect of a photo-excited transient strain field in P3HT and its blends with the electron acceptor [6,6]-phenyl-C61-butyric acid methyl ester (PCBM) obtained using SUEM. We measure the local change of the secondary electron emission as a result of photo-excitation, which gives rise to the photo-induced SUEM image contrast. We observe a ring-shaped spatial profile with both bright and dark contrasts within the photo-illuminated area, indicating a nontrivial spatial distribution of the change in the secondary electron emission yield. The observed spatial profile develops with a rise time of $\sim 300$ ps, beyond which the profile persists up to the limit of the measurement time window (2.7 ns), indicating the absence of spatial diffusion of photocarriers. Instead, we attribute the observed spatial profile to the local modulation of the electronic structure by a photo-induced radial strain that stretches or compresses the polymer chains aligned along the in-plane direction. This conclusion is supported by a finite element simulation of the dynamic elastic response of the polymer film to the photo-induced stress. Due to the low excitation level and the resulting small temperature rise, we propose that the photo-induced stress is primarily originated from the photostriction effect, caused by the coupling of photoexcited charge carriers with the polymer chains, rather than from thermal expansion. Our work demonstrates a promising route towards tailoring opto-electronic properties using transient mechanical strains in conducting polymers, and demonstrates the versatility of SUEM to study photo-physical processes in a broad variety of materials beyond inorganic materials.

\section{Results and Discussion}
\subsection{SUEM Measurement}
\begin{figure}[hbt!]
\centering
\includegraphics[width=\columnwidth,keepaspectratio]{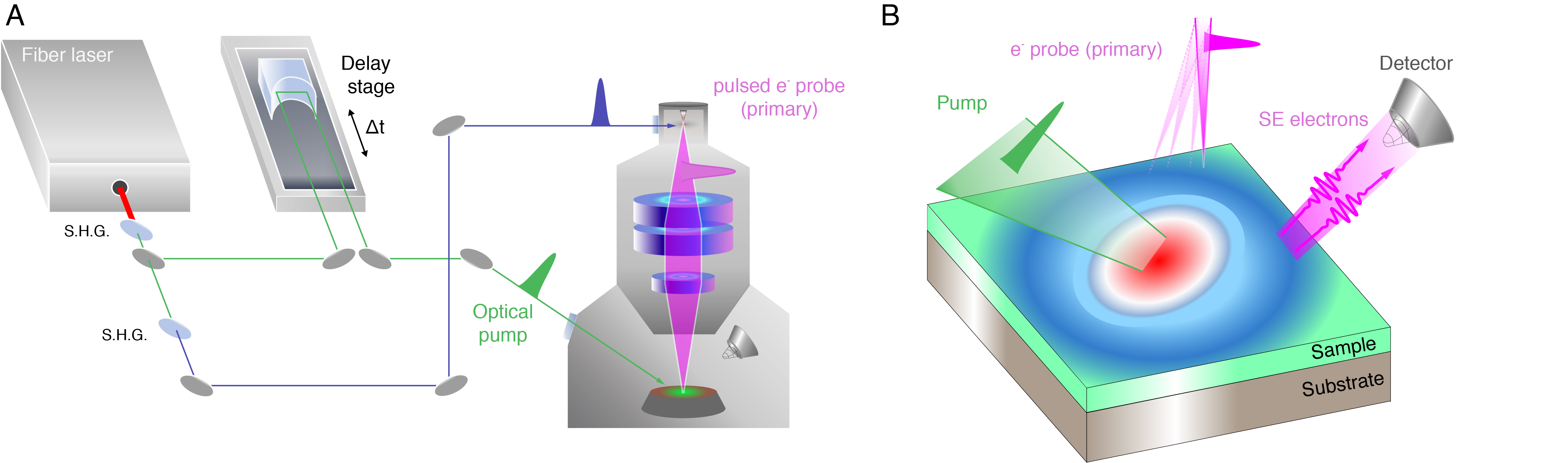}
\caption{\textbf{Schematic Illustration of the Experiment.} (A) Schematic of the SUEM setup. (B) Schematic illustration of the SUEM image formation process. (Inset: optical image of the sample studied in this work (in-plane dimension: $\sim 100$ mm$^2$, thickness of the P3HT: $\sim 60$ nm).}
\label{fig:schematic}
\end{figure}

We employed SUEM to obtain time-resolved images of the photo-induced response in P3HT and P3HT/PCBM thin films deposited on doped silicon substrates. The procedure for sample preparation is identical to that in Ref.~\cite{LimChabinyc_Chemmater} and described in Methods. A schematic of the SUEM setup is shown in Fig.~\ref{fig:schematic}A and the details of the setup are given in Methods. In brief, a fundamental infrared (IR) laser pulse train undergoes harmonic generations to create the visible pump beam (wavelength: 515 nm) and the ultraviolet (UV) photoelectron excitation beam (wavelength: 257 nm). The visible pump beam travels variable distances, yielding a pump-probe time delay from $\sim$ $-$1.3 to 2.7 ns, and is focused onto the specimen to initiate a photo-physical process upon absorption. The UV excitation beam is directed through the column of a scanning electron microscope (SEM) and onto the apex of a cooled Schottky field emission gun, generating electron pulses with sub-picosecond durations via the photoelectric effect. The photo-generated electron pulses (the ``primary" electrons) are accelerated inside the SEM column to 30 keV kinetic energy, and are finely focused to nanometer size through the electron optics in the SEM. Secondary electrons emitted locally from the sample surface upon the impact of the primary electron pulses are collected with a standard Everhart-Thornley detector (ETD). %

Figure~\ref{fig:schematic}B further illustrates the optical excitation and the subsequent detection process. The pump beam optically excites the sample, causing changes in the sample that can alter the secondary electron emission yield after the impact of the primary electron pulses. The local change of the number of emitted secondary electrons is collected and used to form the SUEM contrast images. Time resolution is achieved by controlling the delay time between the optical pump pulses and the electron probe pulses using a mechanical delay stage. Several mechanisms can contribute to the SUEM contrast~\cite{liao2017scanning}, including accumulation of photocarriers~\cite{Bolin_Natnano},  photo-induced modification of the surface potential~\cite{KRONIK_SPV,li2020probing}, and photo-induced topographical changes~\cite{Najafi_P3HT}. By monitoring the space-time evolution of the secondary electron contrast, the relevant photo-physical process on the sample surface can be directly visualized with simultaneous high spatial and temporal resolutions. In this experiment, we believe the SUEM contrast stems from the local modulation of the P3HT electronic structure due to the photo-induced elastic deformation, as will be discussed later.

\begin{figure}[hbt!]
\centering
\includegraphics[width=\columnwidth,keepaspectratio]{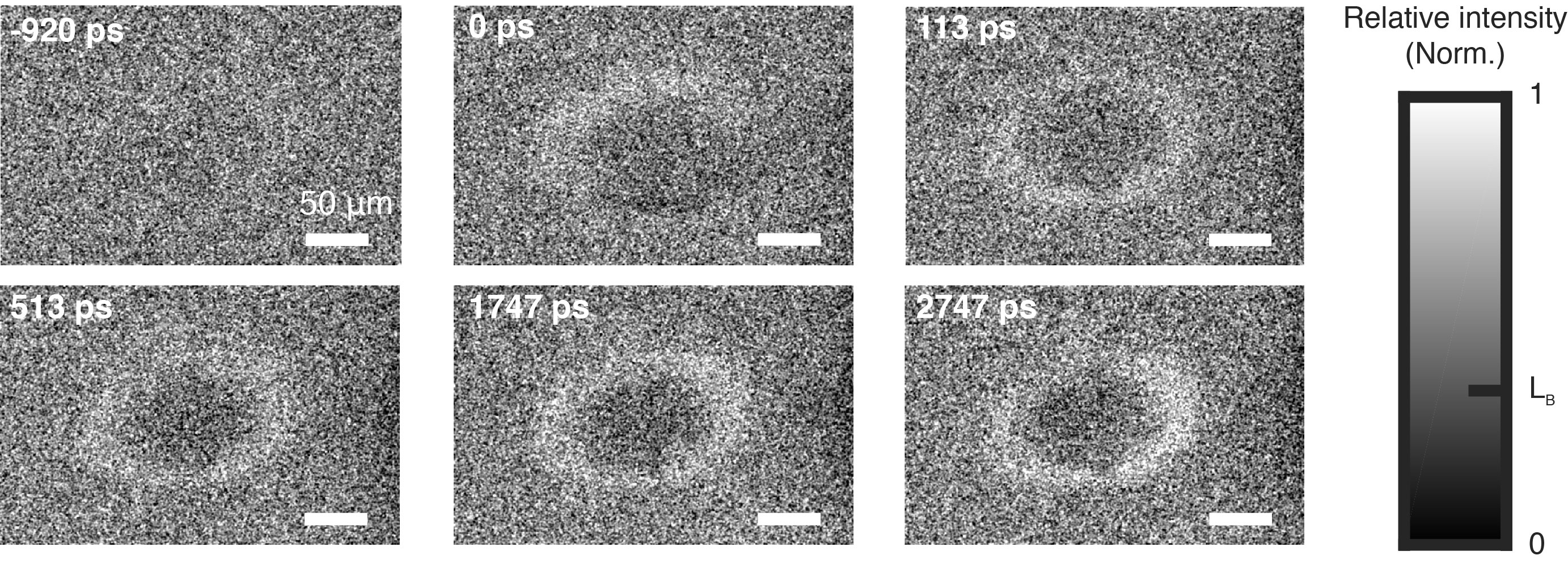}
\caption{\textbf{Representative Time-resolved SUEM Contrast Images of Neat P3HT Films at Several Delay Times.} At each delay time, the measurements were conducted with the electron-beam dwell time of 100 ns and a few thousand scans were averaged to obtain each image. Subsequently, contrast images were formed by subtracting a reference image taken at $-1.3$ ns time delay. The measurements were conducted at multiple locations on the samples with various optical powers (see \SM~Sec. I for additional data). For visual clarity, a spatial Gaussian filter and a linear contrast stretching were used to process the raw images. The color bar indicates the normalized relative intensity distribution at each time delay, in which the $L_B$ labels the average background intensity level.}
\label{fig:diffimg}
\end{figure}

\subsection{Image Processing and Model Analysis}
Figure~\ref{fig:diffimg} shows representative SUEM contrast images of a neat P3HT sample taken at different time delays with an optical pump power of 800 $\mu$W (the corresponding optical fluence is approximately 10 $\mu J$ $cm^{-2}$). At each time delay, a total of 2500 to 7500 scans were averaged to generate one SUEM image with sufficient signal-to-noise ratio (SNR). The dwell time of the electron beam at each pixel was 100 ns. The SUEM contrast images were obtained by subtracting a reference image taken at the pump-probe delay of $\sim - 1.3$ ns when no photo-induced dynamics occur~\cite{Najafi_Science}. The laser repetition rate was set at 1 MHz, corresponding to 1 $\mu$s interval between pulses, to ensure the excited area returns to its initial state before the next pump pulse arrives. The absence of obvious photo-induced contrast in SUEM contrast images taken at negative time delays confirms that the 1 $\mu$s interval is sufficient for this purpose. For visual clarity, the raw images were processed using a spatial low-pass Gaussian filter to remove high-spatial-frequency noises and improve the SNR along with a contrast stretching with a linear intensity mapping. As shown in Fig.~\ref{fig:diffimg}, we observe the generation of an initial spatial intensity contrast immediately after the optical excitation. The horizontal diameter of the profile is $\sim 100$ $\mu m$, comparable to the incident laser spot diameter. A ring-shaped spatial profile with the bright contrast in the outer region and dark contrast in the center is clearly visible, caused by local changes of the secondary electron emission as a result of the photoexcitation \cite{Reimer}. The measured image contrast near the center of the ring is lower than the background level, indicating the photo-induced secondary electron emission yield near the center is lower than the surroundings without photoexcitation. While the intensity of the ring profile evolves with the delay time up to 2.7 ns, there is no obvious change of the shape of the spatial profile. Additional measurements were performed on a P3HT/1 wt\% PCBM blend. The resulting photo-induced dynamics were similar to those in the neat P3HT sample except that a stronger contrast was observed, signaling a stronger photo-response in the P3HT/PCBM blend (See \SM~Sec. II for further details). We also conducted experiments with varying optical pump powers (See more data in \SM~Sec.~I). No qualitative difference was observed when the optical pump fluence was changed below $\sim12$ $\mu J$~$cm^{-2}$, while higher optical pump power led to photo-charging of the sample that would obscure the SUEM contrasts.

\begin{figure}[hbt!]
\centering
\includegraphics[width=1\columnwidth,keepaspectratio]{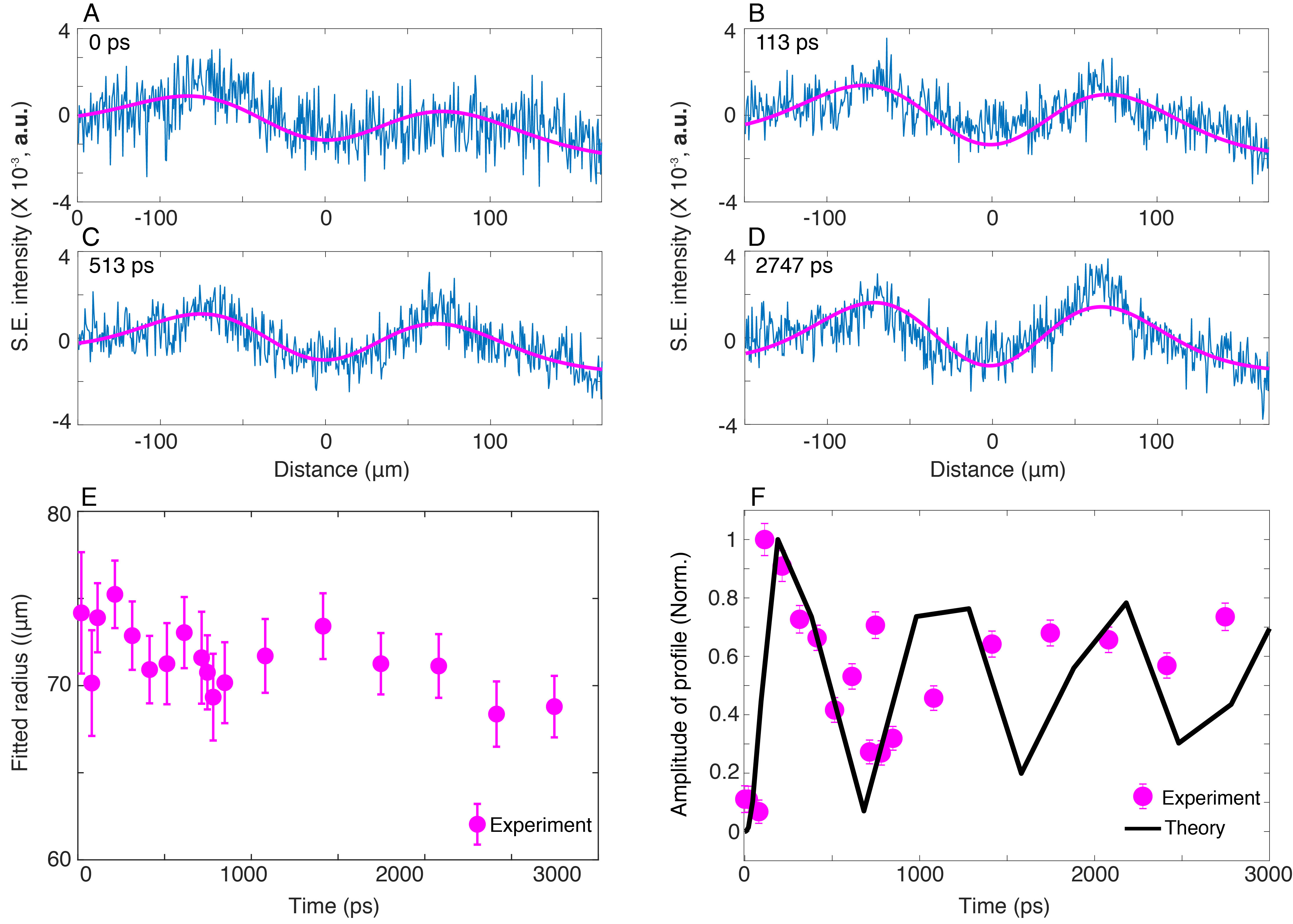}
\caption{\textbf{Quantitative Analysis of the SUEM Contrast Images of Neat P3HT.} (A-D) Measured horizontal intensity line-cut profiles at several delay times (blue lines), along with the best fits using the empirical model described in the text (magenta lines). The horizontal profiles are an average of the intensity along 40 lines near the center of the photo-excited area. (E) Fitted radius of the spatial profiles versus the delay time. The radius corresponds to the parameter $w$ in Eq.~\ref{eq:1}. The error bars were numerically determined with a 95\% confidence interval. (F) Fitted amplitude of the spatial profiles versus the delay time. The error bars were numerically determined with a 95\% confidence interval. Within the experimental uncertainty, the intensity rapidly increases below $\sim 200-300$ ps, above which the magnitude exhibits a periodic oscillation that is subsequently damped beyond 1 ns. The simulated amplitude of the radial strain versus time using a finite element method is given as the solid black line to compare to the experimental results.}
\label{fig:Linecut}
\end{figure}

We quantitatively analyzed the spatial-temporal evolution of the observed photo-induced contrast profile as follows. We first generated horizontal line-cut profiles from the contrast images as given in Figs.~\ref{fig:Linecut}A-D. More precisely, each line-cut profile in Figs.~\ref{fig:Linecut}A-D is an average of the image contrast along 40 horizontal lines near the center of the photo-induced spatial profile. From the line-cut profiles, we observed a dip at the origin, corresponding to the dark contrast in the center region in the contrast images, along with two humps at $\sim \pm 70$ $\mu m$, corresponding to the bright ring region. In passing, we note the asymmetry in the background intensity at distances above $\pm 150$ $\mu m$ from the center, which arises from the fact that the distance from the ETD varies across the sample. This artifact was also observed in previous SUEM measurements~\cite{Bolin_Natnano} and needs to be considered when modeling the observed SUEM contrast.

Next, we used an empirical model to analyze the line-cut profiles in Figs.~\ref{fig:Linecut}A-D. Since the optical pump pulse has a Gaussian spatial intensity distribution, we heuristically considered an empirical function that qualitatively captures the observed spatial profile with the following mathematical form:

\begin{equation}
    f(r) =  a r^2 \exp \left({-r^2}/w^2\right)+mr+b,
    \label{eq:1}
\end{equation}

where $r$ is the radial coordinate, $w$ is the radius of the spatial Gaussian distribution, $m$ and $b$ are parameters used to correct the asymmetric background due to the placement of the ETD, and $a$ is the amplitude of the spatial profile. The model is similar to the expression of the second-order spatial derivative of a Gaussian function. We used this model to fit the experimental line-cut profiles using norm-square minimization. The fitted profiles are given as the solid lines in Figs.~\ref{fig:Linecut}A-D, showing a good agreement with the experimental profiles. The fitted parameter $w$, which is the effective radius of the photo-induced spatial profile, is shown in Fig.~\ref{fig:Linecut}E as a function of the time delay. We observe that $w$ remains relatively constant around the value of $\sim 70$ $\mu$m after the spatial profile emerges upon photo-excitation within the time window considered in this work. In contrast to prior SUEM experiments~\cite{Bolin_Natnano, Bolin_nl,najafi2017super}, where the spatial diffusion of the photogenerated charge carriers was observed, the relative constant value of the radius $w$ in our current work signals that the photocarrier diffusion was not captured within the measurement time window. This is consistent with the low diffusivity of photogenerated excitons and free charge carriers in conducting polymers including P3HT~\cite{Wenkai_APL,shaw2008exciton,Dicker_PRB_2004}. In addtion, we note that the balance of excitons and free charge carriers should be different in the P3HT and P3HT/PCBM blend, with the latter having a higher free charge generation yield proportional to the concentration of PCBM~\cite{savenije2013revealing}. However, both samples have nearly identical qualitative SUEM features, indicating that the observed features are not a result of photocarrier transport. 

Despite the lack of a marked spatial diffusion, we observe nontrivial temporal evolution of the amplitude of the spatial profile, as shown in Fig.~\ref{fig:Linecut}F. Within the uncertainty of the experiment, we see an initial rise of the amplitude on the order of $\sim 300$ ps, followed by an oscillation of the amplitude with a period of $\sim 1000$ ps. The amplitude oscillation implies that the observed spatial profile is related to the dynamic elastic response of the P3HT thin film initiated by a photo-induced stress. In fact, the initial rising time matches the round trip travel time of an acoustic wavefront across the thin film thickness (given the sample thickness $d=60~\rm{nm}$ and the P3HT speed of sound $v=~550$~$\rm{m~s^{-1}}$~\cite{Najafi_P3HT}), which determines the time scale for an elastic strain field to be established after an impulsive excitation.

\begin{figure}[hbt!]
\centering
\includegraphics[width=.9\columnwidth,keepaspectratio]{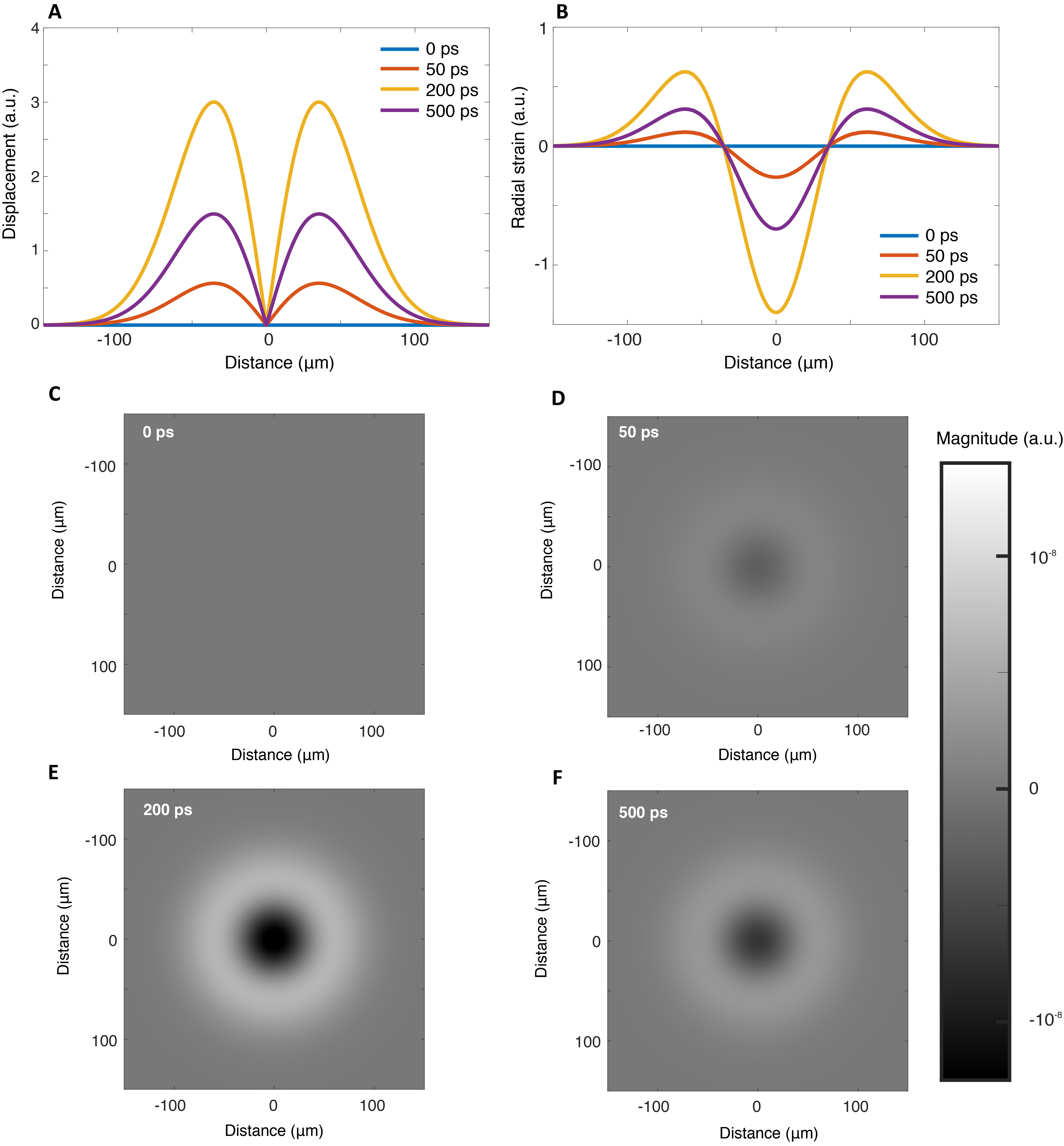}
\caption{\textbf{Calculated Transient Radial Strain after Photoexcitation using a Finite Element Analysis.} (A) Calculated radial displacement versus the radial distance at several time delays. The origin indicates the center of the laser incidence spot (laser spot radius used: 50 $\mu$m). (B) The calculated amplitude of the radial strain versus the radial distance. The radial strain was obtained using the spatial derivative of the profiles in (A). (C-F) The simulated 2D radial strain distribution on the surface of a P3HT thin film at several delay times after photoexcitation. The experimentally observed ring-shaped profiles were reproduced in the simulation. The color bar indicates the amplitude of the strain distribution.}
\label{fig:Theory}
\end{figure}

\subsection{Finite Element Simulation of Elastic Response}
To verify this hypothesis, we simulated the dynamic evolution of the elastic strain field in a P3HT thin film after the photo-excitation using the finite element method (FEM) as implemented in COMSOL Multiphysics. We solved the time-dependent linear elasticity equation with an initial stress field with the same spatial profile as the optical Gaussian pump beam, while imposing a free top surface and a fixed bottom surface of the thin film. Given the axial symmetry of the configuration, two displacement components were solved: the out-of-plane displacement $u_z$ and the radial displacement $u_r$. Although $u_z$ has a larger amplitude than $u_r$ from the simulation, $u_z$ follows a simple Gaussian spatial profile that cannot explain the experimental ring-shaped profile. The radial displacements $u_r$ at different times are shown in Fig.~\ref{fig:Theory}A. The characteristic spatial profile of the radial displacement is due to the zero-displacement boundary condition at the origin in an axial symmetric system. We further calculated the radial strain $\epsilon_{r}=\frac{du_r}{dr}$ that reflects the stretching or compression of the polymer chains along the in-plane direction. The resulting radial strain profile at the surface of the sample is shown in Fig~\ref{fig:Theory}B, in which the positive/negative strain corresponds to a compressive/tensile strain, respectively. We also plot the simulated two-dimensional profiles of the radial strain on the sample surface in Figs~\ref{fig:Theory}C-F. We can see from the simulation that the radial strain profiles qualitatively agree with the experimental profiles shown in Figs.~\ref{fig:Linecut}A-D. Physically, this radial strain profile reflects the tendency of the central region to expand after the photo-excitation while the radial displacement far away from the excited region remains zero, resulting in a compressively strained region near the edge of the photo-excited area. Furthermore, the simulated amplitude of the spatial profile as a function of time is compared to the experimental values in Fig.~\ref{fig:Linecut}F, showing a good agreement before $\sim 1000$ ps. The triangular waveform is characteristic of a thickness resonance excited in a thin film, as observed in prior photoacoustic measurements~\cite{liao2016photo}. However, a discrepancy between the experiment and the simulation can be seen particularly beyond 1000 ps. A probable reason for the discrepancy is that the elastic damping and structural relaxation of the strained polymer film can occur at longer time scales, which were not included in the FEM simulation. 

Several observations are worth a more detailed discussion. First, we note that the secondary electron yield, directly detected by SUEM, is more sensitive to the radial strain than the out-of-plane strain, which has a larger amplitude from the FEM simulation. We attribute this observation to the anisotropic polymer chain alignment in a thin polymer film. It is well established that the chains in P3HT are dominantly aligned along the substrate in thin films~\cite{gurau2007measuring}. Therefore, the in-plane strain should have a direct impact on the backbone while the out-of-plane strain mainly affects the interchain distance. This view is consistent with the analysis by Hess et al.~\cite{hess1993photoexcitations} about the optical properties of P3HT under a hydrostatic pressure. With isotropic applied hydrostatic pressure, a shift in the optical absorption edge  of P3HT was observed suggesting a change in polymer chain conformation; in contrast a change in interchain electronic coupling would modify the shape of the spectrum~\cite{clark2007role}. Second, the observed bright contrast in the ring-shaped region under a compressive strain indicates that a compressive strain on the polymer chains enhances the secondary electron emission, while the dark contrast in the central region under tensile strain implies the opposite effect. Although there is no reported quantum mechanical simulation of the ionization potential of strained P3HT, we postulate the following impact on the crystalline and amorphous regions of the polymer. Given that compressive hydrostatic pressure lowers the optical gap, it suggests that the valence level should be lowered as well leading to increased electron emission.  In the tensile region at these low strains, it is likely that behavior may be dominated by conformational changes in the wider gap amorphous regions leading to a smaller overall change in electronic structure. Third, we further investigated the origin of the photo-induced stress that initiated the dynamic elastic response of the P3HT film. We estimated the temperature rise caused by the absorption of the pump pulse to be within 1 K and calculated the resulting thermal stress due to thermal expansion (see \SM~Sec. IV for further discussion). We found the displacement induced by the thermal stress is exceedingly small and, thus, is unlikely to be responsible for the observed material response. Instead, we suspect the initial stress is directly induced by the generation of photocarriers and the strong coupling between the charge carriers and the polymer chains. This so-called ``photostriction" effect has been reported in inorganic semiconductors~\cite{datskos1998photoinduced}, hybrid perovskites~\cite{zhou2016giant} as well as conducting polymers~\cite{chen2021photostrictive} but has not been investigated in the context of spectroscopic studies. A related ``electrostriction" effect, where a stress is generated by an external electric field, has also been suggested to occur in polymer transistors~\cite{street2008bias}. Our results suggest that, even at moderate photoexcitation levels ($\sim 10^{18}~ \rm{cm}^{-3}$ photocarrier concentration in our experiments), the photostrictive stress along the polymer chains can be significantly higher than the thermal stress in P3HT and P3HT/PCBM blends. In comparison, conventional spectroscopic methods such as optical pump-probe techniques~\cite{kaake2012influence,kanner1990picosecond} measure the average response of the material within the illuminated region and, thus, have not revealed the complex spatial strain profile caused by photo-excitation to date. Furthermore, due to the relatively large optical absorption depth (tens to hundreds of nanometers), conventional optical spectroscopic methods are more sensitive to the photo-induced changes in the bulk, e.g. the thickness change due to the out-of-plane displacement. In contrast, SUEM has superior surface sensitivity due to the small secondary electron escape length (typically few nanometers), and thus, is more sensitive to the radial strain that has not been detected in prior optical measurements.     

\section{Summary}
In summary, we report SUEM measurements of the space-time responses of the semiconducting polymer P3HT after photo-excitation. Combining an empirical model and an FEM simulation, we attribute the observed intriguing ring-shaped spatial profile of the secondary electron emission to the local modulation of the polymer's electronic structure by a photo-induced radial strain. The radial strain compresses or stretches the polymer chains along the in-plane direction, which changes the electronic structure of the polymer due to the electron-lattice coupling. We further estimated the amplitude of the photo-induced stress and concluded that the major contributor to the observed radial strain is the photocarrier generation, namely the photostriction effect, rather than thermal stress. Our results suggest that the photostriction effect can play an important role in the photo-response of conducting polymers, and demonstrates the versatility of the SUEM technique to investigate a wide range of materials beyond inorganic substances.

\section{Methods}

\subsection{Sample Preparation}
First, a lightly Sb-doped silicon wafer (thickness: $\sim$ 530 $\mu$m, crystal orientation $\langle100\rangle$, chlorinated oxide thickness: 200 nm, average resistivity: 0.012 $\Omega\cdot$cm; purchased from International Wafer Service was prepared. The wafer was first wet-cleaned in several ultrasonic baths: soapy water, DI water, acetone, and isopropanol for 5 minutes, respectively, and subsequently dry-cleaned in an oxygen ambient plasma chamber for 5 minutes. The wafer was cleaved to create a square substrate with an in-plane dimension of $\sim100$ $\rm{mm}^2$ using a diamond cutter. Regioregular P3HT (regioregularity: 95\%, product name: SP001 from Merck KGaA) dissolved into chlorobenzene (anhydrous, Sigma Aldrich) (concentration: 10 mg $\rm{mL}^{-1}$), after which the solution was stirred at 350 K to fully dissolve the polymer. The solution was filtered using a 0.45 um polytetrafluoroethylene (PTFE) membrane. Finally, the supported P3HT thin-film was prepared by a two-step spin-coating process (speed: 1000 rpm, acceleration: 500 acc, time: 45 seconds in step 1; speed: 3000 rpm, acceleration: 3000 acc, time: 30 seconds, in step 2). The thickness of the P3HT layer was measured to be $\sim60$ nm using a step profiler (DektakXT Stylus Profilometer by Bruker).
For the present thin film, the H-like aggregate character in the ordered regions of the films was found, which does not exhibit a marked difference upon the addition of the PCBM (1 wt\%), as measured using ultraviolet-visible (UV-VIS) optical characterizations (See \SM~Sec.~III for further details).


\subsection{Scanning Ultrafast Electron Microscope}
 A fundamental infrared (IR) laser pulse train (Clark MXR IMPULSE, wavelength: 1030 nm, pulse duration: 150 fs, repetition rate: 1 MHz) is directed to beta barium borate (BBO) frequency-doubling crystals to create the visible pump beam (wavelength: 515 nm, power: $\sim 800$ $\mu$W) and the ultraviolet (UV) photoelectron excitation beam (wavelength: 343 nm, power: $\sim 15$ mW). The visible pump beam travels variable distances adjusted by a mechanical delay stage (Newport DL600, delay time range: $\sim$ $-$1.3 to 2.7 ns). The UV excitation beam is directed through a window on the column of an SEM (ThermoFisher Quanta 650 FEG) and onto the apex of a cooled Schottky field emission gun (a zirconium-oxide-coated tungsten tip), generating electron pulses with subpicosecond durations via the photoelectric effect. An electron current ($I$) of $\backsimeq 18$, and $80$ pA is used in this experiment, corresponding to $\sim 110$, and $480$ electrons per pulse ($n$). For the results presented in the main text, $I$ of 80 pA was used. The photo-generated electron pulses are accelerated inside the SEM column to 30 keV kinetic energy, and are finely focused to nanometer size through the electron optics in the SEM. During the measurements, the photocathode was refreshed every $\sim 60$ minutes to prevent the fluctuation of the cathode work function. A mechanical coupling system is built to make a rigid connection between the SEM air-suspension system and the optical table hosting the laser and the optical system to minimize the relative vibration that affects the alignment at the photocathode.

\begin{acknowledgments}
We acknowledge helpful discussions with Alexei Maznev. Alejandro Vega-Flick, Alex Ackerman and Yu Li also contributed to the development of the SUEM setup at UCSB. This work is based on research supported by the U.S. Army Research Office under the award number W911NF-19-1-0060 for studying photocarrier dynamics and the U.S. Department of Energy, Office of Basic Energy Sciences, Division of Materials Science and Engineering under the award number DE-SC0019244 for developing SUEM. M.L.C. and S.O. are supported by National Science Foundation, Division of Materials Research under the award number DMR-1808622 for research on thermal and electronic behavior of polymers. 
\end{acknowledgments}

\bibliography{references.bib}

\begin{thebibliography}{55}%
\makeatletter
\providecommand \@ifxundefined [1]{%
 \@ifx{#1\undefined}
}%
\providecommand \@ifnum [1]{%
 \ifnum #1\expandafter \@firstoftwo
 \else \expandafter \@secondoftwo
 \fi
}%
\providecommand \@ifx [1]{%
 \ifx #1\expandafter \@firstoftwo
 \else \expandafter \@secondoftwo
 \fi
}%
\providecommand \natexlab [1]{#1}%
\providecommand \enquote  [1]{``#1''}%
\providecommand \bibnamefont  [1]{#1}%
\providecommand \bibfnamefont [1]{#1}%
\providecommand \citenamefont [1]{#1}%
\providecommand \href@noop [0]{\@secondoftwo}%
\providecommand \href [0]{\begingroup \@sanitize@url \@href}%
\providecommand \@href[1]{\@@startlink{#1}\@@href}%
\providecommand \@@href[1]{\endgroup#1\@@endlink}%
\providecommand \@sanitize@url [0]{\catcode `\\12\catcode `\$12\catcode
  `\&12\catcode `\#12\catcode `\^12\catcode `\_12\catcode `\%12\relax}%
\providecommand \@@startlink[1]{}%
\providecommand \@@endlink[0]{}%
\providecommand \url  [0]{\begingroup\@sanitize@url \@url }%
\providecommand \@url [1]{\endgroup\@href {#1}{\urlprefix }}%
\providecommand \urlprefix  [0]{URL }%
\providecommand \Eprint [0]{\href }%
\providecommand \doibase [0]{https://doi.org/}%
\providecommand \selectlanguage [0]{\@gobble}%
\providecommand \bibinfo  [0]{\@secondoftwo}%
\providecommand \bibfield  [0]{\@secondoftwo}%
\providecommand \translation [1]{[#1]}%
\providecommand \BibitemOpen [0]{}%
\providecommand \bibitemStop [0]{}%
\providecommand \bibitemNoStop [0]{.\EOS\space}%
\providecommand \EOS [0]{\spacefactor3000\relax}%
\providecommand \BibitemShut  [1]{\csname bibitem#1\endcsname}%
\let\auto@bib@innerbib\@empty
\bibitem [{\citenamefont {Yang}\ \emph {et~al.}(2019)\citenamefont {Yang},
  \citenamefont {Mun}, \citenamefont {Kwon}, \citenamefont {Park},
  \citenamefont {Bao},\ and\ \citenamefont {Park}}]{yang2019electronic}%
  \BibitemOpen
  \bibfield  {author} {\bibinfo {author} {\bibfnamefont {J.~C.}\ \bibnamefont
  {Yang}}, \bibinfo {author} {\bibfnamefont {J.}~\bibnamefont {Mun}}, \bibinfo
  {author} {\bibfnamefont {S.~Y.}\ \bibnamefont {Kwon}}, \bibinfo {author}
  {\bibfnamefont {S.}~\bibnamefont {Park}}, \bibinfo {author} {\bibfnamefont
  {Z.}~\bibnamefont {Bao}}, and\ \bibinfo {author} {\bibfnamefont
  {S.}~\bibnamefont {Park}},\ }\bibfield  {title} {\bibinfo {title} {Electronic
  skin: Recent progress and future prospects for skin-attachable devices for
  health monitoring, robotics, and prosthetics},\ }\href
  {https://doi.org/https://doi.org/10.1002/adma.201904765} {\bibfield
  {journal} {\bibinfo  {journal} {Advanced Materials}\ }\textbf {\bibinfo
  {volume} {31}},\ \bibinfo {pages} {1904765} (\bibinfo {year} {2019})},\
  \Eprint
  {https://arxiv.org/abs/https://onlinelibrary.wiley.com/doi/pdf/10.1002/adma.201904765}
  {https://onlinelibrary.wiley.com/doi/pdf/10.1002/adma.201904765} \BibitemShut
  {NoStop}%
\bibitem [{\citenamefont {Someya}\ \emph {et~al.}(2017)\citenamefont {Someya},
  \citenamefont {Bauer},\ and\ \citenamefont
  {Kaltenbrunner}}]{someya2017imperceptible}%
  \BibitemOpen
  \bibfield  {author} {\bibinfo {author} {\bibfnamefont {T.}~\bibnamefont
  {Someya}}, \bibinfo {author} {\bibfnamefont {S.}~\bibnamefont {Bauer}}, and\
  \bibinfo {author} {\bibfnamefont {M.}~\bibnamefont {Kaltenbrunner}},\
  }\bibfield  {title} {\bibinfo {title} {Imperceptible organic electronics},\
  }\href {https://doi.org/10.1557/mrs.2017.1} {\bibfield  {journal} {\bibinfo
  {journal} {MRS Bulletin}\ }\textbf {\bibinfo {volume} {42}},\ \bibinfo
  {pages} {124–130} (\bibinfo {year} {2017})}\BibitemShut {NoStop}%
\bibitem [{\citenamefont {Yan}\ \emph {et~al.}(2018)\citenamefont {Yan},
  \citenamefont {Barlow}, \citenamefont {Wang}, \citenamefont {Yan},
  \citenamefont {Jen}, \citenamefont {Marder},\ and\ \citenamefont
  {Zhan}}]{yan2018non}%
  \BibitemOpen
  \bibfield  {author} {\bibinfo {author} {\bibfnamefont {C.}~\bibnamefont
  {Yan}}, \bibinfo {author} {\bibfnamefont {S.}~\bibnamefont {Barlow}},
  \bibinfo {author} {\bibfnamefont {Z.}~\bibnamefont {Wang}}, \bibinfo {author}
  {\bibfnamefont {H.}~\bibnamefont {Yan}}, \bibinfo {author} {\bibfnamefont
  {A.~K.-Y.}\ \bibnamefont {Jen}}, \bibinfo {author} {\bibfnamefont {S.~R.}\
  \bibnamefont {Marder}}, and\ \bibinfo {author} {\bibfnamefont
  {X.}~\bibnamefont {Zhan}},\ }\bibfield  {title} {\bibinfo {title}
  {Non-fullerene acceptors for organic solar cells},\ }\href
  {https://doi.org/10.1038/natrevmats.2018.3} {\bibfield  {journal} {\bibinfo
  {journal} {Nature Reviews Materials}\ }\textbf {\bibinfo {volume} {3}},\
  \bibinfo {pages} {18003} (\bibinfo {year} {2018})}\BibitemShut {NoStop}%
\bibitem [{\citenamefont {Inal}\ \emph {et~al.}(2018)\citenamefont {Inal},
  \citenamefont {Rivnay}, \citenamefont {Suiu}, \citenamefont {Malliaras},\
  and\ \citenamefont {McCulloch}}]{inal2018conjugated}%
  \BibitemOpen
  \bibfield  {author} {\bibinfo {author} {\bibfnamefont {S.}~\bibnamefont
  {Inal}}, \bibinfo {author} {\bibfnamefont {J.}~\bibnamefont {Rivnay}},
  \bibinfo {author} {\bibfnamefont {A.-O.}\ \bibnamefont {Suiu}}, \bibinfo
  {author} {\bibfnamefont {G.~G.}\ \bibnamefont {Malliaras}}, and\ \bibinfo
  {author} {\bibfnamefont {I.}~\bibnamefont {McCulloch}},\ }\bibfield  {title}
  {\bibinfo {title} {Conjugated polymers in bioelectronics},\ }\href
  {https://doi.org/10.1021/acs.accounts.7b00624} {\bibfield  {journal}
  {\bibinfo  {journal} {Accounts of Chemical Research}\ }\textbf {\bibinfo
  {volume} {51}},\ \bibinfo {pages} {1368} (\bibinfo {year} {2018})},\ \Eprint
  {https://arxiv.org/abs/https://doi.org/10.1021/acs.accounts.7b00624}
  {https://doi.org/10.1021/acs.accounts.7b00624} \BibitemShut {NoStop}%
\bibitem [{\citenamefont {Mei\:}\ and\ \citenamefont
  {Bao}(2014)}]{MeiandZBao_COM}%
  \BibitemOpen
  \bibfield  {author} {\bibinfo {author} {\bibfnamefont {J.}~\bibnamefont
  {Mei\:}}and\ \bibinfo {author} {\bibfnamefont {Z.}~\bibnamefont {Bao}},\
  }\bibfield  {title} {\bibinfo {title} {Side chain engineering in
  solution-processable conjugated polymers},\ }\href
  {https://doi.org/10.1021/cm4020805} {\bibfield  {journal} {\bibinfo
  {journal} {Chemistry of Materials}\ }\textbf {\bibinfo {volume} {26}},\
  \bibinfo {pages} {604} (\bibinfo {year} {2014})},\ \Eprint
  {https://arxiv.org/abs/https://doi.org/10.1021/cm4020805}
  {https://doi.org/10.1021/cm4020805} \BibitemShut {NoStop}%
\bibitem [{\citenamefont {Paterson}\ \emph {et~al.}(2018)\citenamefont
  {Paterson}, \citenamefont {Singh}, \citenamefont {Fallon}, \citenamefont
  {Hodsden}, \citenamefont {Han}, \citenamefont {Schroeder}, \citenamefont
  {Bronstein}, \citenamefont {Heeney}, \citenamefont {McCulloch},\ and\
  \citenamefont {Anthopoulos}}]{paterson2018recent}%
  \BibitemOpen
  \bibfield  {author} {\bibinfo {author} {\bibfnamefont {A.~F.}\ \bibnamefont
  {Paterson}}, \bibinfo {author} {\bibfnamefont {S.}~\bibnamefont {Singh}},
  \bibinfo {author} {\bibfnamefont {K.~J.}\ \bibnamefont {Fallon}}, \bibinfo
  {author} {\bibfnamefont {T.}~\bibnamefont {Hodsden}}, \bibinfo {author}
  {\bibfnamefont {Y.}~\bibnamefont {Han}}, \bibinfo {author} {\bibfnamefont
  {B.~C.}\ \bibnamefont {Schroeder}}, \bibinfo {author} {\bibfnamefont
  {H.}~\bibnamefont {Bronstein}}, \bibinfo {author} {\bibfnamefont
  {M.}~\bibnamefont {Heeney}}, \bibinfo {author} {\bibfnamefont
  {I.}~\bibnamefont {McCulloch}}, and\ \bibinfo {author} {\bibfnamefont
  {T.~D.}\ \bibnamefont {Anthopoulos}},\ }\bibfield  {title} {\bibinfo {title}
  {Recent progress in high-mobility organic transistors: A reality check},\
  }\href {https://doi.org/https://doi.org/10.1002/adma.201801079} {\bibfield
  {journal} {\bibinfo  {journal} {Advanced Materials}\ }\textbf {\bibinfo
  {volume} {30}},\ \bibinfo {pages} {1801079} (\bibinfo {year} {2018})},\
  \Eprint
  {https://arxiv.org/abs/https://onlinelibrary.wiley.com/doi/pdf/10.1002/adma.201801079}
  {https://onlinelibrary.wiley.com/doi/pdf/10.1002/adma.201801079} \BibitemShut
  {NoStop}%
\bibitem [{\citenamefont {Ghosh\:}\ and\ \citenamefont
  {Spano}(2020)}]{ghosh2020excitons}%
  \BibitemOpen
  \bibfield  {author} {\bibinfo {author} {\bibfnamefont {R.}~\bibnamefont
  {Ghosh\:}}and\ \bibinfo {author} {\bibfnamefont {F.~C.}\ \bibnamefont
  {Spano}},\ }\bibfield  {title} {\bibinfo {title} {Excitons and polarons in
  organic materials},\ }\href {https://doi.org/10.1021/acs.accounts.0c00349}
  {\bibfield  {journal} {\bibinfo  {journal} {Accounts of Chemical Research}\
  }\textbf {\bibinfo {volume} {53}},\ \bibinfo {pages} {2201} (\bibinfo {year}
  {2020})},\ \bibinfo {note} {pMID: 33035054},\ \Eprint
  {https://arxiv.org/abs/https://doi.org/10.1021/acs.accounts.0c00349}
  {https://doi.org/10.1021/acs.accounts.0c00349} \BibitemShut {NoStop}%
\bibitem [{\citenamefont {Xie}\ \emph {et~al.}(2018)\citenamefont {Xie},
  \citenamefont {Colby},\ and\ \citenamefont {Gomez}}]{xie2018connecting}%
  \BibitemOpen
  \bibfield  {author} {\bibinfo {author} {\bibfnamefont {R.}~\bibnamefont
  {Xie}}, \bibinfo {author} {\bibfnamefont {R.~H.}\ \bibnamefont {Colby}}, and\
  \bibinfo {author} {\bibfnamefont {E.~D.}\ \bibnamefont {Gomez}},\ }\bibfield
  {title} {\bibinfo {title} {Connecting the mechanical and conductive
  properties of conjugated polymers},\ }\href
  {https://doi.org/https://doi.org/10.1002/aelm.201700356} {\bibfield
  {journal} {\bibinfo  {journal} {Advanced Electronic Materials}\ }\textbf
  {\bibinfo {volume} {4}},\ \bibinfo {pages} {1700356} (\bibinfo {year}
  {2018})},\ \Eprint
  {https://arxiv.org/abs/https://onlinelibrary.wiley.com/doi/pdf/10.1002/aelm.201700356}
  {https://onlinelibrary.wiley.com/doi/pdf/10.1002/aelm.201700356} \BibitemShut
  {NoStop}%
\bibitem [{\citenamefont {Root}\ \emph {et~al.}(2017)\citenamefont {Root},
  \citenamefont {Savagatrup}, \citenamefont {Printz}, \citenamefont
  {Rodriquez},\ and\ \citenamefont {Lipomi}}]{root2017mechanical}%
  \BibitemOpen
  \bibfield  {author} {\bibinfo {author} {\bibfnamefont {S.~E.}\ \bibnamefont
  {Root}}, \bibinfo {author} {\bibfnamefont {S.}~\bibnamefont {Savagatrup}},
  \bibinfo {author} {\bibfnamefont {A.~D.}\ \bibnamefont {Printz}}, \bibinfo
  {author} {\bibfnamefont {D.}~\bibnamefont {Rodriquez}}, and\ \bibinfo
  {author} {\bibfnamefont {D.~J.}\ \bibnamefont {Lipomi}},\ }\bibfield  {title}
  {\bibinfo {title} {Mechanical properties of organic semiconductors for
  stretchable, highly flexible, and mechanically robust electronics},\ }\href
  {https://doi.org/10.1021/acs.chemrev.7b00003} {\bibfield  {journal} {\bibinfo
   {journal} {Chemical Reviews}\ }\textbf {\bibinfo {volume} {117}},\ \bibinfo
  {pages} {6467} (\bibinfo {year} {2017})},\ \bibinfo {note} {pMID: 28343389},\
  \Eprint {https://arxiv.org/abs/https://doi.org/10.1021/acs.chemrev.7b00003}
  {https://doi.org/10.1021/acs.chemrev.7b00003} \BibitemShut {NoStop}%
\bibitem [{\citenamefont {Poelking\:}\ and\ \citenamefont
  {Andrienko}(2013)}]{poelking2013effect}%
  \BibitemOpen
  \bibfield  {author} {\bibinfo {author} {\bibfnamefont {C.}~\bibnamefont
  {Poelking\:}}and\ \bibinfo {author} {\bibfnamefont {D.}~\bibnamefont
  {Andrienko}},\ }\bibfield  {title} {\bibinfo {title} {Effect of polymorphism,
  regioregularity and paracrystallinity on charge transport in
  poly(3-hexylthiophene) [p3ht] nanofibers},\ }\href
  {https://doi.org/10.1021/ma4015966} {\bibfield  {journal} {\bibinfo
  {journal} {Macromolecules}\ }\textbf {\bibinfo {volume} {46}},\ \bibinfo
  {pages} {8941} (\bibinfo {year} {2013})},\ \Eprint
  {https://arxiv.org/abs/https://doi.org/10.1021/ma4015966}
  {https://doi.org/10.1021/ma4015966} \BibitemShut {NoStop}%
\bibitem [{\citenamefont {Menichetti}\ \emph {et~al.}(2017)\citenamefont
  {Menichetti}, \citenamefont {Colle},\ and\ \citenamefont
  {Grosso}}]{menichetti2017strain}%
  \BibitemOpen
  \bibfield  {author} {\bibinfo {author} {\bibfnamefont {G.}~\bibnamefont
  {Menichetti}}, \bibinfo {author} {\bibfnamefont {R.}~\bibnamefont {Colle}},
  and\ \bibinfo {author} {\bibfnamefont {G.}~\bibnamefont {Grosso}},\
  }\bibfield  {title} {\bibinfo {title} {Strain modulation of band offsets at
  the {PCBM/P3HT} heterointerface},\ }\href
  {https://doi.org/10.1021/acs.jpcc.7b02717} {\bibfield  {journal} {\bibinfo
  {journal} {The Journal of Physical Chemistry C}\ }\textbf {\bibinfo {volume}
  {121}},\ \bibinfo {pages} {13707} (\bibinfo {year} {2017})},\ \Eprint
  {https://arxiv.org/abs/https://doi.org/10.1021/acs.jpcc.7b02717}
  {https://doi.org/10.1021/acs.jpcc.7b02717} \BibitemShut {NoStop}%
\bibitem [{\citenamefont {Qian}\ \emph {et~al.}(2016)\citenamefont {Qian},
  \citenamefont {Zhang}, \citenamefont {Xie}, \citenamefont {Qi}, \citenamefont
  {Chandran}, \citenamefont {Chen},\ and\ \citenamefont
  {Huang}}]{qian2016stretchable}%
  \BibitemOpen
  \bibfield  {author} {\bibinfo {author} {\bibfnamefont {Y.}~\bibnamefont
  {Qian}}, \bibinfo {author} {\bibfnamefont {X.}~\bibnamefont {Zhang}},
  \bibinfo {author} {\bibfnamefont {L.}~\bibnamefont {Xie}}, \bibinfo {author}
  {\bibfnamefont {D.}~\bibnamefont {Qi}}, \bibinfo {author} {\bibfnamefont
  {B.~K.}\ \bibnamefont {Chandran}}, \bibinfo {author} {\bibfnamefont
  {X.}~\bibnamefont {Chen}}, and\ \bibinfo {author} {\bibfnamefont
  {W.}~\bibnamefont {Huang}},\ }\bibfield  {title} {\bibinfo {title}
  {Stretchable organic semiconductor devices},\ }\href
  {https://doi.org/https://doi.org/10.1002/adma.201601278} {\bibfield
  {journal} {\bibinfo  {journal} {Advanced Materials}\ }\textbf {\bibinfo
  {volume} {28}},\ \bibinfo {pages} {9243} (\bibinfo {year} {2016})},\ \Eprint
  {https://arxiv.org/abs/https://onlinelibrary.wiley.com/doi/pdf/10.1002/adma.201601278}
  {https://onlinelibrary.wiley.com/doi/pdf/10.1002/adma.201601278} \BibitemShut
  {NoStop}%
\bibitem [{\citenamefont {Hess}\ \emph {et~al.}(1993)\citenamefont {Hess},
  \citenamefont {Kanner},\ and\ \citenamefont
  {Vardeny}}]{hess1993photoexcitations}%
  \BibitemOpen
  \bibfield  {author} {\bibinfo {author} {\bibfnamefont {B.~C.}\ \bibnamefont
  {Hess}}, \bibinfo {author} {\bibfnamefont {G.~S.}\ \bibnamefont {Kanner}},
  and\ \bibinfo {author} {\bibfnamefont {Z.}~\bibnamefont {Vardeny}},\
  }\bibfield  {title} {\bibinfo {title} {Photoexcitations in polythiophene at
  high pressure},\ }\href {https://doi.org/10.1103/PhysRevB.47.1407} {\bibfield
   {journal} {\bibinfo  {journal} {Phys. Rev. B}\ }\textbf {\bibinfo {volume}
  {47}},\ \bibinfo {pages} {1407} (\bibinfo {year} {1993})}\BibitemShut
  {NoStop}%
\bibitem [{\citenamefont {Schmidtke}\ \emph {et~al.}(2007)\citenamefont
  {Schmidtke}, \citenamefont {Kim}, \citenamefont {Gierschner}, \citenamefont
  {Silva},\ and\ \citenamefont {Friend}}]{schmidtke2007optical}%
  \BibitemOpen
  \bibfield  {author} {\bibinfo {author} {\bibfnamefont {J.~P.}\ \bibnamefont
  {Schmidtke}}, \bibinfo {author} {\bibfnamefont {J.-S.}\ \bibnamefont {Kim}},
  \bibinfo {author} {\bibfnamefont {J.}~\bibnamefont {Gierschner}}, \bibinfo
  {author} {\bibfnamefont {C.}~\bibnamefont {Silva}}, and\ \bibinfo {author}
  {\bibfnamefont {R.~H.}\ \bibnamefont {Friend}},\ }\bibfield  {title}
  {\bibinfo {title} {Optical spectroscopy of a polyfluorene copolymer at high
  pressure: Intra- and intermolecular interactions},\ }\href
  {https://doi.org/10.1103/PhysRevLett.99.167401} {\bibfield  {journal}
  {\bibinfo  {journal} {Phys. Rev. Lett.}\ }\textbf {\bibinfo {volume} {99}},\
  \bibinfo {pages} {167401} (\bibinfo {year} {2007})}\BibitemShut {NoStop}%
\bibitem [{\citenamefont {Chen}\ \emph {et~al.}(2019)\citenamefont {Chen},
  \citenamefont {Rastak}, \citenamefont {Wang}, \citenamefont {Yan},
  \citenamefont {Feig}, \citenamefont {Liu}, \citenamefont {Jiang},
  \citenamefont {Chen}, \citenamefont {Lian}, \citenamefont {Molina-Lopez},
  \citenamefont {Jin}, \citenamefont {Cui}, \citenamefont {Chung},
  \citenamefont {Pop}, \citenamefont {Linder},\ and\ \citenamefont
  {Bao}}]{ChenandZBao_Mat}%
  \BibitemOpen
  \bibfield  {author} {\bibinfo {author} {\bibfnamefont {G.}~\bibnamefont
  {Chen}}, \bibinfo {author} {\bibfnamefont {R.}~\bibnamefont {Rastak}},
  \bibinfo {author} {\bibfnamefont {Y.}~\bibnamefont {Wang}}, \bibinfo {author}
  {\bibfnamefont {H.}~\bibnamefont {Yan}}, \bibinfo {author} {\bibfnamefont
  {V.}~\bibnamefont {Feig}}, \bibinfo {author} {\bibfnamefont {Y.}~\bibnamefont
  {Liu}}, \bibinfo {author} {\bibfnamefont {Y.}~\bibnamefont {Jiang}}, \bibinfo
  {author} {\bibfnamefont {S.}~\bibnamefont {Chen}}, \bibinfo {author}
  {\bibfnamefont {F.}~\bibnamefont {Lian}}, \bibinfo {author} {\bibfnamefont
  {F.}~\bibnamefont {Molina-Lopez}}, \bibinfo {author} {\bibfnamefont
  {L.}~\bibnamefont {Jin}}, \bibinfo {author} {\bibfnamefont {K.}~\bibnamefont
  {Cui}}, \bibinfo {author} {\bibfnamefont {J.~W.}\ \bibnamefont {Chung}},
  \bibinfo {author} {\bibfnamefont {E.}~\bibnamefont {Pop}}, \bibinfo {author}
  {\bibfnamefont {C.}~\bibnamefont {Linder}}, and\ \bibinfo {author}
  {\bibfnamefont {Z.}~\bibnamefont {Bao}},\ }\bibfield  {title} {\bibinfo
  {title} {Strain- and strain-rate-invariant conductance in a stretchable and
  compressible {3D} conducting polymer foam},\ }\href
  {https://doi.org/https://doi.org/10.1016/j.matt.2019.03.011} {\bibfield
  {journal} {\bibinfo  {journal} {Matter}\ }\textbf {\bibinfo {volume} {1}},\
  \bibinfo {pages} {205} (\bibinfo {year} {2019})}\BibitemShut {NoStop}%
\bibitem [{\citenamefont {Vijay}\ \emph {et~al.}(2011)\citenamefont {Vijay},
  \citenamefont {Rao},\ and\ \citenamefont {Narayan}}]{Vijay_JAP_2011}%
  \BibitemOpen
  \bibfield  {author} {\bibinfo {author} {\bibfnamefont {V.}~\bibnamefont
  {Vijay}}, \bibinfo {author} {\bibfnamefont {A.~D.}\ \bibnamefont {Rao}}, and\
  \bibinfo {author} {\bibfnamefont {K.~S.}\ \bibnamefont {Narayan}},\
  }\bibfield  {title} {\bibinfo {title} {In situ studies of strain dependent
  transport properties of conducting polymers on elastomeric substrates},\
  }\href {https://doi.org/10.1063/1.3580514} {\bibfield  {journal} {\bibinfo
  {journal} {Journal of Applied Physics}\ }\textbf {\bibinfo {volume} {109}},\
  \bibinfo {pages} {084525} (\bibinfo {year} {2011})},\ \Eprint
  {https://arxiv.org/abs/https://doi.org/10.1063/1.3580514}
  {https://doi.org/10.1063/1.3580514} \BibitemShut {NoStop}%
\bibitem [{\citenamefont {Oh}\ \emph {et~al.}(2016)\citenamefont {Oh},
  \citenamefont {Rondeau-Gagn{\'e}}, \citenamefont {Chiu}, \citenamefont
  {Chortos}, \citenamefont {Lissel}, \citenamefont {Wang}, \citenamefont
  {Schroeder}, \citenamefont {Kurosawa}, \citenamefont {Lopez}, \citenamefont
  {Katsumata}, \citenamefont {Xu}, \citenamefont {Zhu}, \citenamefont {Gu},
  \citenamefont {Bae}, \citenamefont {Kim}, \citenamefont {Jin}, \citenamefont
  {Chung}, \citenamefont {Tok},\ and\ \citenamefont
  {Bao}}]{OhandZBao_Nature_2016}%
  \BibitemOpen
  \bibfield  {author} {\bibinfo {author} {\bibfnamefont {J.~Y.}\ \bibnamefont
  {Oh}}, \bibinfo {author} {\bibfnamefont {S.}~\bibnamefont
  {Rondeau-Gagn{\'e}}}, \bibinfo {author} {\bibfnamefont {Y.-C.}\ \bibnamefont
  {Chiu}}, \bibinfo {author} {\bibfnamefont {A.}~\bibnamefont {Chortos}},
  \bibinfo {author} {\bibfnamefont {F.}~\bibnamefont {Lissel}}, \bibinfo
  {author} {\bibfnamefont {G.-J.~N.}\ \bibnamefont {Wang}}, \bibinfo {author}
  {\bibfnamefont {B.~C.}\ \bibnamefont {Schroeder}}, \bibinfo {author}
  {\bibfnamefont {T.}~\bibnamefont {Kurosawa}}, \bibinfo {author}
  {\bibfnamefont {J.}~\bibnamefont {Lopez}}, \bibinfo {author} {\bibfnamefont
  {T.}~\bibnamefont {Katsumata}}, \bibinfo {author} {\bibfnamefont
  {J.}~\bibnamefont {Xu}}, \bibinfo {author} {\bibfnamefont {C.}~\bibnamefont
  {Zhu}}, \bibinfo {author} {\bibfnamefont {X.}~\bibnamefont {Gu}}, \bibinfo
  {author} {\bibfnamefont {W.-G.}\ \bibnamefont {Bae}}, \bibinfo {author}
  {\bibfnamefont {Y.}~\bibnamefont {Kim}}, \bibinfo {author} {\bibfnamefont
  {L.}~\bibnamefont {Jin}}, \bibinfo {author} {\bibfnamefont {J.~W.}\
  \bibnamefont {Chung}}, \bibinfo {author} {\bibfnamefont {J.~B.-H.}\
  \bibnamefont {Tok}}, and\ \bibinfo {author} {\bibfnamefont {Z.}~\bibnamefont
  {Bao}},\ }\bibfield  {title} {\bibinfo {title} {Intrinsically stretchable and
  healable semiconducting polymer for organic transistors},\ }\href
  {https://doi.org/10.1038/nature20102} {\bibfield  {journal} {\bibinfo
  {journal} {Nature}\ }\textbf {\bibinfo {volume} {539}},\ \bibinfo {pages}
  {411} (\bibinfo {year} {2016})}\BibitemShut {NoStop}%
\bibitem [{\citenamefont {Sahabudeen}\ \emph {et~al.}(2016)\citenamefont
  {Sahabudeen}, \citenamefont {Qi}, \citenamefont {Glatz}, \citenamefont
  {Tranca}, \citenamefont {Dong}, \citenamefont {Hou}, \citenamefont {Zhang},
  \citenamefont {Kuttner}, \citenamefont {Lehnert}, \citenamefont {Seifert},
  \citenamefont {Kaiser}, \citenamefont {Fery}, \citenamefont {Zheng},\ and\
  \citenamefont {Feng}}]{Sahabudeen_Natcomm}%
  \BibitemOpen
  \bibfield  {author} {\bibinfo {author} {\bibfnamefont {H.}~\bibnamefont
  {Sahabudeen}}, \bibinfo {author} {\bibfnamefont {H.}~\bibnamefont {Qi}},
  \bibinfo {author} {\bibfnamefont {B.~A.}\ \bibnamefont {Glatz}}, \bibinfo
  {author} {\bibfnamefont {D.}~\bibnamefont {Tranca}}, \bibinfo {author}
  {\bibfnamefont {R.}~\bibnamefont {Dong}}, \bibinfo {author} {\bibfnamefont
  {Y.}~\bibnamefont {Hou}}, \bibinfo {author} {\bibfnamefont {T.}~\bibnamefont
  {Zhang}}, \bibinfo {author} {\bibfnamefont {C.}~\bibnamefont {Kuttner}},
  \bibinfo {author} {\bibfnamefont {T.}~\bibnamefont {Lehnert}}, \bibinfo
  {author} {\bibfnamefont {G.}~\bibnamefont {Seifert}}, \bibinfo {author}
  {\bibfnamefont {U.}~\bibnamefont {Kaiser}}, \bibinfo {author} {\bibfnamefont
  {A.}~\bibnamefont {Fery}}, \bibinfo {author} {\bibfnamefont {Z.}~\bibnamefont
  {Zheng}}, and\ \bibinfo {author} {\bibfnamefont {X.}~\bibnamefont {Feng}},\
  }\bibfield  {title} {\bibinfo {title} {Wafer-sized multifunctional
  polyimine-based two-dimensional conjugated polymers with high mechanical
  stiffness},\ }\href {https://doi.org/10.1038/ncomms13461} {\bibfield
  {journal} {\bibinfo  {journal} {Nature Communications}\ }\textbf {\bibinfo
  {volume} {7}},\ \bibinfo {pages} {13461} (\bibinfo {year}
  {2016})}\BibitemShut {NoStop}%
\bibitem [{\citenamefont {Gasperini}\ \emph {et~al.}(2019)\citenamefont
  {Gasperini}, \citenamefont {Wang}, \citenamefont {Molina-Lopez},
  \citenamefont {Wu}, \citenamefont {Lopez}, \citenamefont {Xu}, \citenamefont
  {Luo}, \citenamefont {Zhou}, \citenamefont {Xue}, \citenamefont {Tok},\ and\
  \citenamefont {Bao}}]{GasperiniandZBao_MAMO}%
  \BibitemOpen
  \bibfield  {author} {\bibinfo {author} {\bibfnamefont {A.}~\bibnamefont
  {Gasperini}}, \bibinfo {author} {\bibfnamefont {G.-J.~N.}\ \bibnamefont
  {Wang}}, \bibinfo {author} {\bibfnamefont {F.}~\bibnamefont {Molina-Lopez}},
  \bibinfo {author} {\bibfnamefont {H.-C.}\ \bibnamefont {Wu}}, \bibinfo
  {author} {\bibfnamefont {J.}~\bibnamefont {Lopez}}, \bibinfo {author}
  {\bibfnamefont {J.}~\bibnamefont {Xu}}, \bibinfo {author} {\bibfnamefont
  {S.}~\bibnamefont {Luo}}, \bibinfo {author} {\bibfnamefont {D.}~\bibnamefont
  {Zhou}}, \bibinfo {author} {\bibfnamefont {G.}~\bibnamefont {Xue}}, \bibinfo
  {author} {\bibfnamefont {J.~B.-H.}\ \bibnamefont {Tok}}, and\ \bibinfo
  {author} {\bibfnamefont {Z.}~\bibnamefont {Bao}},\ }\bibfield  {title}
  {\bibinfo {title} {Characterization of hydrogen bonding formation and
  breaking in semiconducting polymers under mechanical strain},\ }\href
  {https://doi.org/10.1021/acs.macromol.9b00145} {\bibfield  {journal}
  {\bibinfo  {journal} {Macromolecules}\ }\textbf {\bibinfo {volume} {52}},\
  \bibinfo {pages} {2476} (\bibinfo {year} {2019})},\ \Eprint
  {https://arxiv.org/abs/https://doi.org/10.1021/acs.macromol.9b00145}
  {https://doi.org/10.1021/acs.macromol.9b00145} \BibitemShut {NoStop}%
\bibitem [{\citenamefont {Turner}\ \emph {et~al.}(2011)\citenamefont {Turner},
  \citenamefont {Pingel}, \citenamefont {Steyrleuthner}, \citenamefont
  {Crossland}, \citenamefont {Ludwigs},\ and\ \citenamefont
  {Neher}}]{Turner_AdvFuncMat}%
  \BibitemOpen
  \bibfield  {author} {\bibinfo {author} {\bibfnamefont {S.~T.}\ \bibnamefont
  {Turner}}, \bibinfo {author} {\bibfnamefont {P.}~\bibnamefont {Pingel}},
  \bibinfo {author} {\bibfnamefont {R.}~\bibnamefont {Steyrleuthner}}, \bibinfo
  {author} {\bibfnamefont {E.~J.~W.}\ \bibnamefont {Crossland}}, \bibinfo
  {author} {\bibfnamefont {S.}~\bibnamefont {Ludwigs}}, and\ \bibinfo {author}
  {\bibfnamefont {D.}~\bibnamefont {Neher}},\ }\bibfield  {title} {\bibinfo
  {title} {Quantitative analysis of bulk heterojunction films using linear
  absorption spectroscopy and solar cell performance},\ }\href
  {https://doi.org/https://doi.org/10.1002/adfm.201101583} {\bibfield
  {journal} {\bibinfo  {journal} {Advanced Functional Materials}\ }\textbf
  {\bibinfo {volume} {21}},\ \bibinfo {pages} {4640} (\bibinfo {year}
  {2011})},\ \Eprint
  {https://arxiv.org/abs/https://onlinelibrary.wiley.com/doi/pdf/10.1002/adfm.201101583}
  {https://onlinelibrary.wiley.com/doi/pdf/10.1002/adfm.201101583} \BibitemShut
  {NoStop}%
\bibitem [{\citenamefont {Awartani}\ \emph {et~al.}(2013)\citenamefont
  {Awartani}, \citenamefont {Lemanski}, \citenamefont {Ro}, \citenamefont
  {Richter}, \citenamefont {DeLongchamp},\ and\ \citenamefont
  {O'Connor}}]{Awartani_AdvEnMat}%
  \BibitemOpen
  \bibfield  {author} {\bibinfo {author} {\bibfnamefont {O.}~\bibnamefont
  {Awartani}}, \bibinfo {author} {\bibfnamefont {B.~I.}\ \bibnamefont
  {Lemanski}}, \bibinfo {author} {\bibfnamefont {H.~W.}\ \bibnamefont {Ro}},
  \bibinfo {author} {\bibfnamefont {L.~J.}\ \bibnamefont {Richter}}, \bibinfo
  {author} {\bibfnamefont {D.~M.}\ \bibnamefont {DeLongchamp}}, and\ \bibinfo
  {author} {\bibfnamefont {B.~T.}\ \bibnamefont {O'Connor}},\ }\bibfield
  {title} {\bibinfo {title} {Correlating stiffness, ductility, and morphology
  of polymer:fullerene films for solar cell applications},\ }\href
  {https://doi.org/https://doi.org/10.1002/aenm.201200595} {\bibfield
  {journal} {\bibinfo  {journal} {Advanced Energy Materials}\ }\textbf
  {\bibinfo {volume} {3}},\ \bibinfo {pages} {399} (\bibinfo {year} {2013})},\
  \Eprint
  {https://arxiv.org/abs/https://onlinelibrary.wiley.com/doi/pdf/10.1002/aenm.201200595}
  {https://onlinelibrary.wiley.com/doi/pdf/10.1002/aenm.201200595} \BibitemShut
  {NoStop}%
\bibitem [{\citenamefont {Moon}\ \emph
  {et~al.}(2020{\natexlab{a}})\citenamefont {Moon}, \citenamefont {Grosso},
  \citenamefont {Chakraborty}, \citenamefont {Peng}, \citenamefont {Taniguchi},
  \citenamefont {Watanabe},\ and\ \citenamefont {Englund}}]{Moon_nl}%
  \BibitemOpen
  \bibfield  {author} {\bibinfo {author} {\bibfnamefont {H.}~\bibnamefont
  {Moon}}, \bibinfo {author} {\bibfnamefont {G.}~\bibnamefont {Grosso}},
  \bibinfo {author} {\bibfnamefont {C.}~\bibnamefont {Chakraborty}}, \bibinfo
  {author} {\bibfnamefont {C.}~\bibnamefont {Peng}}, \bibinfo {author}
  {\bibfnamefont {T.}~\bibnamefont {Taniguchi}}, \bibinfo {author}
  {\bibfnamefont {K.}~\bibnamefont {Watanabe}}, and\ \bibinfo {author}
  {\bibfnamefont {D.}~\bibnamefont {Englund}},\ }\bibfield  {title} {\bibinfo
  {title} {Dynamic exciton funneling by local strain control in a monolayer
  semiconductor},\ }\href {https://doi.org/10.1021/acs.nanolett.0c02757}
  {\bibfield  {journal} {\bibinfo  {journal} {Nano Letters}\ }\textbf {\bibinfo
  {volume} {20}},\ \bibinfo {pages} {6791} (\bibinfo {year}
  {2020}{\natexlab{a}})},\ \bibinfo {note} {pMID: 32790415},\ \Eprint
  {https://arxiv.org/abs/https://doi.org/10.1021/acs.nanolett.0c02757}
  {https://doi.org/10.1021/acs.nanolett.0c02757} \BibitemShut {NoStop}%
\bibitem [{\citenamefont {Moon}\ \emph
  {et~al.}(2020{\natexlab{b}})\citenamefont {Moon}, \citenamefont {Bersin},
  \citenamefont {Chakraborty}, \citenamefont {Lu}, \citenamefont {Grosso},
  \citenamefont {Kong},\ and\ \citenamefont {Englund}}]{Moon_ACSPho}%
  \BibitemOpen
  \bibfield  {author} {\bibinfo {author} {\bibfnamefont {H.}~\bibnamefont
  {Moon}}, \bibinfo {author} {\bibfnamefont {E.}~\bibnamefont {Bersin}},
  \bibinfo {author} {\bibfnamefont {C.}~\bibnamefont {Chakraborty}}, \bibinfo
  {author} {\bibfnamefont {A.-Y.}\ \bibnamefont {Lu}}, \bibinfo {author}
  {\bibfnamefont {G.}~\bibnamefont {Grosso}}, \bibinfo {author} {\bibfnamefont
  {J.}~\bibnamefont {Kong}}, and\ \bibinfo {author} {\bibfnamefont
  {D.}~\bibnamefont {Englund}},\ }\bibfield  {title} {\bibinfo {title}
  {Strain-correlated localized exciton energy in atomically thin
  semiconductors},\ }\href {https://doi.org/10.1021/acsphotonics.0c00626}
  {\bibfield  {journal} {\bibinfo  {journal} {ACS Photonics}\ }\textbf
  {\bibinfo {volume} {7}},\ \bibinfo {pages} {1135} (\bibinfo {year}
  {2020}{\natexlab{b}})},\ \Eprint
  {https://arxiv.org/abs/https://doi.org/10.1021/acsphotonics.0c00626}
  {https://doi.org/10.1021/acsphotonics.0c00626} \BibitemShut {NoStop}%
\bibitem [{\citenamefont {Sheng}\ \emph {et~al.}(2007)\citenamefont {Sheng},
  \citenamefont {Tong}, \citenamefont {Singh},\ and\ \citenamefont
  {Vardeny}}]{ShengVardeny_PRB}%
  \BibitemOpen
  \bibfield  {author} {\bibinfo {author} {\bibfnamefont {C.-X.}\ \bibnamefont
  {Sheng}}, \bibinfo {author} {\bibfnamefont {M.}~\bibnamefont {Tong}},
  \bibinfo {author} {\bibfnamefont {S.}~\bibnamefont {Singh}}, and\ \bibinfo
  {author} {\bibfnamefont {Z.~V.}\ \bibnamefont {Vardeny}},\ }\bibfield
  {title} {\bibinfo {title} {Experimental determination of the charge/neutral
  branching ratio $\ensuremath{\eta}$ in the photoexcitation of
  $\ensuremath{\pi}$-conjugated polymers by broadband ultrafast spectroscopy},\
  }\href {https://doi.org/10.1103/PhysRevB.75.085206} {\bibfield  {journal}
  {\bibinfo  {journal} {Phys. Rev. B}\ }\textbf {\bibinfo {volume} {75}},\
  \bibinfo {pages} {085206} (\bibinfo {year} {2007})}\BibitemShut {NoStop}%
\bibitem [{\citenamefont {Piris}\ \emph {et~al.}(2009)\citenamefont {Piris},
  \citenamefont {Dykstra}, \citenamefont {Bakulin}, \citenamefont {Loosdrecht},
  \citenamefont {Knulst}, \citenamefont {Trinh}, \citenamefont {Schins},\ and\
  \citenamefont {Siebbeles}}]{Piris_JPCC}%
  \BibitemOpen
  \bibfield  {author} {\bibinfo {author} {\bibfnamefont {J.}~\bibnamefont
  {Piris}}, \bibinfo {author} {\bibfnamefont {T.~E.}\ \bibnamefont {Dykstra}},
  \bibinfo {author} {\bibfnamefont {A.~A.}\ \bibnamefont {Bakulin}}, \bibinfo
  {author} {\bibfnamefont {P.~H.~v.}\ \bibnamefont {Loosdrecht}}, \bibinfo
  {author} {\bibfnamefont {W.}~\bibnamefont {Knulst}}, \bibinfo {author}
  {\bibfnamefont {M.~T.}\ \bibnamefont {Trinh}}, \bibinfo {author}
  {\bibfnamefont {J.~M.}\ \bibnamefont {Schins}}, and\ \bibinfo {author}
  {\bibfnamefont {L.~D.}\ \bibnamefont {Siebbeles}},\ }\bibfield  {title}
  {\bibinfo {title} {Photogeneration and ultrafast dynamics of excitons and
  charges in {P3HT/PCBM} blends},\ }\href {https://doi.org/10.1021/jp904229q}
  {\bibfield  {journal} {\bibinfo  {journal} {The Journal of Physical Chemistry
  C}\ }\textbf {\bibinfo {volume} {113}},\ \bibinfo {pages} {14500} (\bibinfo
  {year} {2009})},\ \Eprint
  {https://arxiv.org/abs/https://doi.org/10.1021/jp904229q}
  {https://doi.org/10.1021/jp904229q} \BibitemShut {NoStop}%
\bibitem [{\citenamefont {Albert-Seifried}\ and\ \citenamefont
  {Friend}(2011)}]{albert2011measurement}%
  \BibitemOpen
  \bibfield  {author} {\bibinfo {author} {\bibfnamefont {S.}~\bibnamefont
  {Albert-Seifried}}and\ \bibinfo {author} {\bibfnamefont {R.~H.}\ \bibnamefont
  {Friend}},\ }\bibfield  {title} {\bibinfo {title} {Measurement of thermal
  modulation of optical absorption in pump-probe spectroscopy of semiconducting
  polymers},\ }\href {https://doi.org/10.1063/1.3595340} {\bibfield  {journal}
  {\bibinfo  {journal} {Applied Physics Letters}\ }\textbf {\bibinfo {volume}
  {98}},\ \bibinfo {pages} {223304} (\bibinfo {year} {2011})},\ \Eprint
  {https://arxiv.org/abs/https://doi.org/10.1063/1.3595340}
  {https://doi.org/10.1063/1.3595340} \BibitemShut {NoStop}%
\bibitem [{\citenamefont {Dicker}\ \emph {et~al.}(2004)\citenamefont {Dicker},
  \citenamefont {de~Haas}, \citenamefont {Siebbeles},\ and\ \citenamefont
  {Warman}}]{Dicker_PRB_2004}%
  \BibitemOpen
  \bibfield  {author} {\bibinfo {author} {\bibfnamefont {G.}~\bibnamefont
  {Dicker}}, \bibinfo {author} {\bibfnamefont {M.~P.}\ \bibnamefont {de~Haas}},
  \bibinfo {author} {\bibfnamefont {L.~D.}\ \bibnamefont {Siebbeles}}, and\
  \bibinfo {author} {\bibfnamefont {J.~M.}\ \bibnamefont {Warman}},\ }\bibfield
   {title} {\bibinfo {title} {Electrodeless time-resolved microwave
  conductivity study of charge-carrier photogeneration in regioregular
  poly(3-hexylthiophene) thin films},\ }\href
  {https://doi.org/10.1103/PhysRevB.70.045203} {\bibfield  {journal} {\bibinfo
  {journal} {Phys. Rev. B}\ }\textbf {\bibinfo {volume} {70}},\ \bibinfo
  {pages} {045203} (\bibinfo {year} {2004})}\BibitemShut {NoStop}%
\bibitem [{\citenamefont {Ouyang}\ \emph {et~al.}(2020)\citenamefont {Ouyang},
  \citenamefont {Li}, \citenamefont {Yurash}, \citenamefont {Schopp},
  \citenamefont {Vega-Flick}, \citenamefont {Brus}, \citenamefont {Nguyen},\
  and\ \citenamefont {Liao}}]{Wenkai_APL}%
  \BibitemOpen
  \bibfield  {author} {\bibinfo {author} {\bibfnamefont {W.}~\bibnamefont
  {Ouyang}}, \bibinfo {author} {\bibfnamefont {Y.}~\bibnamefont {Li}}, \bibinfo
  {author} {\bibfnamefont {B.}~\bibnamefont {Yurash}}, \bibinfo {author}
  {\bibfnamefont {N.}~\bibnamefont {Schopp}}, \bibinfo {author} {\bibfnamefont
  {A.}~\bibnamefont {Vega-Flick}}, \bibinfo {author} {\bibfnamefont
  {V.}~\bibnamefont {Brus}}, \bibinfo {author} {\bibfnamefont {T.-Q.}\
  \bibnamefont {Nguyen}}, and\ \bibinfo {author} {\bibfnamefont
  {B.}~\bibnamefont {Liao}},\ }\bibfield  {title} {\bibinfo {title} {Transient
  grating spectroscopy of photocarrier dynamics in semiconducting polymer thin
  films},\ }\href {https://doi.org/10.1063/5.0034773} {\bibfield  {journal}
  {\bibinfo  {journal} {Applied Physics Letters}\ }\textbf {\bibinfo {volume}
  {117}},\ \bibinfo {pages} {253302} (\bibinfo {year} {2020})},\ \Eprint
  {https://arxiv.org/abs/https://doi.org/10.1063/5.0034773}
  {https://doi.org/10.1063/5.0034773} \BibitemShut {NoStop}%
\bibitem [{\citenamefont {Yang}\ \emph {et~al.}(2010)\citenamefont {Yang},
  \citenamefont {Mohammed},\ and\ \citenamefont {Zewail}}]{Jerry_PNAS}%
  \BibitemOpen
  \bibfield  {author} {\bibinfo {author} {\bibfnamefont {D.-S.}\ \bibnamefont
  {Yang}}, \bibinfo {author} {\bibfnamefont {O.~F.}\ \bibnamefont {Mohammed}},
  and\ \bibinfo {author} {\bibfnamefont {A.~H.}\ \bibnamefont {Zewail}},\
  }\bibfield  {title} {\bibinfo {title} {Scanning ultrafast electron
  microscopy},\ }\href {https://doi.org/10.1073/pnas.1009321107} {\bibfield
  {journal} {\bibinfo  {journal} {Proceedings of the National Academy of
  Sciences}\ }\textbf {\bibinfo {volume} {107}},\ \bibinfo {pages} {14993}
  (\bibinfo {year} {2010})},\ \Eprint
  {https://arxiv.org/abs/https://www.pnas.org/content/107/34/14993.full.pdf}
  {https://www.pnas.org/content/107/34/14993.full.pdf} \BibitemShut {NoStop}%
\bibitem [{\citenamefont {Liao\:}\ and\ \citenamefont
  {Najafi}(2017)}]{liao2017scanning}%
  \BibitemOpen
  \bibfield  {author} {\bibinfo {author} {\bibfnamefont {B.}~\bibnamefont
  {Liao\:}}and\ \bibinfo {author} {\bibfnamefont {E.}~\bibnamefont {Najafi}},\
  }\bibfield  {title} {\bibinfo {title} {Scanning ultrafast electron
  microscopy: A novel technique to probe photocarrier dynamics with high
  spatial and temporal resolutions},\ }\href
  {https://doi.org/https://doi.org/10.1016/j.mtphys.2017.07.003} {\bibfield
  {journal} {\bibinfo  {journal} {Materials Today Physics}\ }\textbf {\bibinfo
  {volume} {2}},\ \bibinfo {pages} {46} (\bibinfo {year} {2017})}\BibitemShut
  {NoStop}%
\bibitem [{\citenamefont {Celio}\ \emph {et~al.}(2016)\citenamefont {Celio},
  \citenamefont {Chandler}, \citenamefont {Talin},\ and\ \citenamefont
  {Michael}}]{KimSandia}%
  \BibitemOpen
  \bibfield  {author} {\bibinfo {author} {\bibfnamefont {K.~C.}\ \bibnamefont
  {Celio}}, \bibinfo {author} {\bibfnamefont {D.~W.}\ \bibnamefont {Chandler}},
  \bibinfo {author} {\bibfnamefont {A.~A.}\ \bibnamefont {Talin}}, and\
  \bibinfo {author} {\bibfnamefont {J.}~\bibnamefont {Michael}},\ }\href@noop
  {} {\bibinfo {title} {Development of scanning ultrafast electron microscope
  capability}} (\bibinfo {year} {2016}),\ \bibinfo {note} {\textsc{S}andia
  National Laboratory Report, Livermore, California}\BibitemShut {NoStop}%
\bibitem [{\citenamefont {Bose}\ \emph {et~al.}(2016)\citenamefont {Bose},
  \citenamefont {Bera}, \citenamefont {Parida}, \citenamefont {Adhikari},
  \citenamefont {Shaheen}, \citenamefont {Alarousu}, \citenamefont {Sun},
  \citenamefont {Wu}, \citenamefont {Bakr},\ and\ \citenamefont
  {Mohammed}}]{BoseandOmar_nl}%
  \BibitemOpen
  \bibfield  {author} {\bibinfo {author} {\bibfnamefont {R.}~\bibnamefont
  {Bose}}, \bibinfo {author} {\bibfnamefont {A.}~\bibnamefont {Bera}}, \bibinfo
  {author} {\bibfnamefont {M.~R.}\ \bibnamefont {Parida}}, \bibinfo {author}
  {\bibfnamefont {A.}~\bibnamefont {Adhikari}}, \bibinfo {author}
  {\bibfnamefont {B.~S.}\ \bibnamefont {Shaheen}}, \bibinfo {author}
  {\bibfnamefont {E.}~\bibnamefont {Alarousu}}, \bibinfo {author}
  {\bibfnamefont {J.}~\bibnamefont {Sun}}, \bibinfo {author} {\bibfnamefont
  {T.}~\bibnamefont {Wu}}, \bibinfo {author} {\bibfnamefont {O.~M.}\
  \bibnamefont {Bakr}}, and\ \bibinfo {author} {\bibfnamefont {O.~F.}\
  \bibnamefont {Mohammed}},\ }\bibfield  {title} {\bibinfo {title} {Real-space
  mapping of surface trap states in {CIGSe} nanocrystals using {4D} electron
  microscopy},\ }\href {https://doi.org/10.1021/acs.nanolett.6b01553}
  {\bibfield  {journal} {\bibinfo  {journal} {Nano Letters}\ }\textbf {\bibinfo
  {volume} {16}},\ \bibinfo {pages} {4417} (\bibinfo {year} {2016})},\ \bibinfo
  {note} {pMID: 27228321},\ \Eprint
  {https://arxiv.org/abs/https://doi.org/10.1021/acs.nanolett.6b01553}
  {https://doi.org/10.1021/acs.nanolett.6b01553} \BibitemShut {NoStop}%
\bibitem [{\citenamefont {Zani}\ \emph {et~al.}(2018)\citenamefont {Zani},
  \citenamefont {Sala}, \citenamefont {Irde}, \citenamefont {Pietralunga},
  \citenamefont {Manzoni}, \citenamefont {Cerullo}, \citenamefont {Lanzani},\
  and\ \citenamefont {Tagliaferri}}]{Zani_SUEM}%
  \BibitemOpen
  \bibfield  {author} {\bibinfo {author} {\bibfnamefont {M.}~\bibnamefont
  {Zani}}, \bibinfo {author} {\bibfnamefont {V.}~\bibnamefont {Sala}}, \bibinfo
  {author} {\bibfnamefont {G.}~\bibnamefont {Irde}}, \bibinfo {author}
  {\bibfnamefont {S.~M.}\ \bibnamefont {Pietralunga}}, \bibinfo {author}
  {\bibfnamefont {C.}~\bibnamefont {Manzoni}}, \bibinfo {author} {\bibfnamefont
  {G.}~\bibnamefont {Cerullo}}, \bibinfo {author} {\bibfnamefont
  {G.}~\bibnamefont {Lanzani}}, and\ \bibinfo {author} {\bibfnamefont
  {A.}~\bibnamefont {Tagliaferri}},\ }\bibfield  {title} {\bibinfo {title}
  {Charge dynamics in aluminum oxide thin film studied by ultrafast scanning
  electron microscopy},\ }\href
  {https://doi.org/https://doi.org/10.1016/j.ultramic.2018.01.010} {\bibfield
  {journal} {\bibinfo  {journal} {Ultramicroscopy}\ }\textbf {\bibinfo {volume}
  {187}},\ \bibinfo {pages} {93} (\bibinfo {year} {2018})}\BibitemShut
  {NoStop}%
\bibitem [{\citenamefont {Najafi}\ \emph {et~al.}(2015)\citenamefont {Najafi},
  \citenamefont {Scarborough}, \citenamefont {Tang},\ and\ \citenamefont
  {Zewail}}]{Najafi_Science}%
  \BibitemOpen
  \bibfield  {author} {\bibinfo {author} {\bibfnamefont {E.}~\bibnamefont
  {Najafi}}, \bibinfo {author} {\bibfnamefont {T.~D.}\ \bibnamefont
  {Scarborough}}, \bibinfo {author} {\bibfnamefont {J.}~\bibnamefont {Tang}},
  and\ \bibinfo {author} {\bibfnamefont {A.}~\bibnamefont {Zewail}},\
  }\bibfield  {title} {\bibinfo {title} {Four-dimensional imaging of carrier
  interface dynamics in p-n junctions},\ }\href
  {https://science.sciencemag.org/content/347/6218/164} {\bibfield  {journal}
  {\bibinfo  {journal} {Science}\ }\textbf {\bibinfo {volume} {347}},\ \bibinfo
  {pages} {164} (\bibinfo {year} {2015})},\ \Eprint
  {https://arxiv.org/abs/https://science.sciencemag.org/content/347/6218/164.full.pdf}
  {https://science.sciencemag.org/content/347/6218/164.full.pdf} \BibitemShut
  {NoStop}%
\bibitem [{\citenamefont {Wong}\ \emph {et~al.}(2021)\citenamefont {Wong},
  \citenamefont {Davoyan}, \citenamefont {Liao}, \citenamefont {Krayev},
  \citenamefont {Jo}, \citenamefont {Rotenberg}, \citenamefont {Bostwick},
  \citenamefont {Jozwiak}, \citenamefont {Jariwala}, \citenamefont {Zewail},\
  and\ \citenamefont {Atwater}}]{wong2021spatiotemporal}%
  \BibitemOpen
  \bibfield  {author} {\bibinfo {author} {\bibfnamefont {J.}~\bibnamefont
  {Wong}}, \bibinfo {author} {\bibfnamefont {A.}~\bibnamefont {Davoyan}},
  \bibinfo {author} {\bibfnamefont {B.}~\bibnamefont {Liao}}, \bibinfo {author}
  {\bibfnamefont {A.}~\bibnamefont {Krayev}}, \bibinfo {author} {\bibfnamefont
  {K.}~\bibnamefont {Jo}}, \bibinfo {author} {\bibfnamefont {E.}~\bibnamefont
  {Rotenberg}}, \bibinfo {author} {\bibfnamefont {A.}~\bibnamefont {Bostwick}},
  \bibinfo {author} {\bibfnamefont {C.~M.}\ \bibnamefont {Jozwiak}}, \bibinfo
  {author} {\bibfnamefont {D.}~\bibnamefont {Jariwala}}, \bibinfo {author}
  {\bibfnamefont {A.~H.}\ \bibnamefont {Zewail}}, and\ \bibinfo {author}
  {\bibfnamefont {H.~A.}\ \bibnamefont {Atwater}},\ }\bibfield  {title}
  {\bibinfo {title} {Spatiotemporal imaging of thickness-induced band-bending
  junctions},\ }\href {https://doi.org/10.1021/acs.nanolett.1c01481} {\bibfield
   {journal} {\bibinfo  {journal} {Nano Letters}\ }\textbf {\bibinfo {volume}
  {21}},\ \bibinfo {pages} {5745} (\bibinfo {year} {2021})},\ \bibinfo {note}
  {pMID: 34152777},\ \Eprint
  {https://arxiv.org/abs/https://doi.org/10.1021/acs.nanolett.1c01481}
  {https://doi.org/10.1021/acs.nanolett.1c01481} \BibitemShut {NoStop}%
\bibitem [{\citenamefont {Liao}\ \emph
  {et~al.}(2017{\natexlab{a}})\citenamefont {Liao}, \citenamefont {Najafi},
  \citenamefont {Li}, \citenamefont {Minnich},\ and\ \citenamefont
  {Zewail}}]{Bolin_Natnano}%
  \BibitemOpen
  \bibfield  {author} {\bibinfo {author} {\bibfnamefont {B.}~\bibnamefont
  {Liao}}, \bibinfo {author} {\bibfnamefont {E.}~\bibnamefont {Najafi}},
  \bibinfo {author} {\bibfnamefont {H.}~\bibnamefont {Li}}, \bibinfo {author}
  {\bibfnamefont {A.~J.}\ \bibnamefont {Minnich}}, and\ \bibinfo {author}
  {\bibfnamefont {A.~H.}\ \bibnamefont {Zewail}},\ }\bibfield  {title}
  {\bibinfo {title} {Photo-excited hot carrier dynamics in hydrogenated
  amorphous silicon imaged by {4D} electron microscopy},\ }\href
  {https://doi.org/10.1038/nnano.2017.124} {\bibfield  {journal} {\bibinfo
  {journal} {Nature Nanotechnology}\ }\textbf {\bibinfo {volume} {12}},\
  \bibinfo {pages} {871} (\bibinfo {year} {2017}{\natexlab{a}})}\BibitemShut
  {NoStop}%
\bibitem [{\citenamefont {Najafi}\ \emph {et~al.}(2017)\citenamefont {Najafi},
  \citenamefont {Ivanov}, \citenamefont {Zewail},\ and\ \citenamefont
  {Bernardi}}]{najafi2017super}%
  \BibitemOpen
  \bibfield  {author} {\bibinfo {author} {\bibfnamefont {E.}~\bibnamefont
  {Najafi}}, \bibinfo {author} {\bibfnamefont {V.}~\bibnamefont {Ivanov}},
  \bibinfo {author} {\bibfnamefont {A.}~\bibnamefont {Zewail}}, and\ \bibinfo
  {author} {\bibfnamefont {M.}~\bibnamefont {Bernardi}},\ }\bibfield  {title}
  {\bibinfo {title} {Super-diffusion of excited carriers in semiconductors},\
  }\href {https://doi.org/10.1038/ncomms15177} {\bibfield  {journal} {\bibinfo
  {journal} {Nature Communications}\ }\textbf {\bibinfo {volume} {8}},\
  \bibinfo {pages} {15177} (\bibinfo {year} {2017})}\BibitemShut {NoStop}%
\bibitem [{\citenamefont {Liao}\ \emph
  {et~al.}(2017{\natexlab{b}})\citenamefont {Liao}, \citenamefont {Zhao},
  \citenamefont {Najafi}, \citenamefont {Yan}, \citenamefont {Tian},
  \citenamefont {Tice}, \citenamefont {Minnich}, \citenamefont {Wang},\ and\
  \citenamefont {Zewail}}]{Bolin_nl}%
  \BibitemOpen
  \bibfield  {author} {\bibinfo {author} {\bibfnamefont {B.}~\bibnamefont
  {Liao}}, \bibinfo {author} {\bibfnamefont {H.}~\bibnamefont {Zhao}}, \bibinfo
  {author} {\bibfnamefont {E.}~\bibnamefont {Najafi}}, \bibinfo {author}
  {\bibfnamefont {X.}~\bibnamefont {Yan}}, \bibinfo {author} {\bibfnamefont
  {H.}~\bibnamefont {Tian}}, \bibinfo {author} {\bibfnamefont {J.}~\bibnamefont
  {Tice}}, \bibinfo {author} {\bibfnamefont {A.~J.}\ \bibnamefont {Minnich}},
  \bibinfo {author} {\bibfnamefont {H.}~\bibnamefont {Wang}}, and\ \bibinfo
  {author} {\bibfnamefont {A.~H.}\ \bibnamefont {Zewail}},\ }\bibfield  {title}
  {\bibinfo {title} {Spatial-temporal imaging of anisotropic photocarrier
  dynamics in black phosphorus},\ }\href
  {https://doi.org/10.1021/acs.nanolett.7b00897} {\bibfield  {journal}
  {\bibinfo  {journal} {Nano Letters}\ }\textbf {\bibinfo {volume} {17}},\
  \bibinfo {pages} {3675} (\bibinfo {year} {2017}{\natexlab{b}})},\ \bibinfo
  {note} {pMID: 28505461},\ \Eprint
  {https://arxiv.org/abs/https://doi.org/10.1021/acs.nanolett.7b00897}
  {https://doi.org/10.1021/acs.nanolett.7b00897} \BibitemShut {NoStop}%
\bibitem [{\citenamefont {Najafi}\ \emph {et~al.}(2021)\citenamefont {Najafi},
  \citenamefont {Jafari},\ and\ \citenamefont {Liao}}]{najafi2021carrier}%
  \BibitemOpen
  \bibfield  {author} {\bibinfo {author} {\bibfnamefont {E.}~\bibnamefont
  {Najafi}}, \bibinfo {author} {\bibfnamefont {A.}~\bibnamefont {Jafari}}, and\
  \bibinfo {author} {\bibfnamefont {B.}~\bibnamefont {Liao}},\ }\bibfield
  {title} {\bibinfo {title} {Carrier density oscillation in the photoexcited
  semiconductor},\ }\href {https://doi.org/10.1088/1361-6463/abd1a4} {\bibfield
   {journal} {\bibinfo  {journal} {Journal of Physics D: Applied Physics}\
  }\textbf {\bibinfo {volume} {54}},\ \bibinfo {pages} {125102} (\bibinfo
  {year} {2021})}\BibitemShut {NoStop}%
\bibitem [{\citenamefont {Najafi}\ \emph {et~al.}(2018)\citenamefont {Najafi},
  \citenamefont {Liao}, \citenamefont {Scarborough},\ and\ \citenamefont
  {Zewail}}]{Najafi_P3HT}%
  \BibitemOpen
  \bibfield  {author} {\bibinfo {author} {\bibfnamefont {E.}~\bibnamefont
  {Najafi}}, \bibinfo {author} {\bibfnamefont {B.}~\bibnamefont {Liao}},
  \bibinfo {author} {\bibfnamefont {T.}~\bibnamefont {Scarborough}}, and\
  \bibinfo {author} {\bibfnamefont {A.}~\bibnamefont {Zewail}},\ }\bibfield
  {title} {\bibinfo {title} {Imaging surface acoustic wave dynamics in
  semiconducting polymers by scanning ultrafast electron microscopy},\ }\href
  {https://doi.org/https://doi.org/10.1016/j.ultramic.2017.08.011} {\bibfield
  {journal} {\bibinfo  {journal} {Ultramicroscopy}\ }\textbf {\bibinfo {volume}
  {184}},\ \bibinfo {pages} {46} (\bibinfo {year} {2018})}\BibitemShut
  {NoStop}%
\bibitem [{\citenamefont {Lim}\ \emph {et~al.}(2018)\citenamefont {Lim},
  \citenamefont {Peterson}, \citenamefont {Su},\ and\ \citenamefont
  {Chabinyc}}]{LimChabinyc_Chemmater}%
  \BibitemOpen
  \bibfield  {author} {\bibinfo {author} {\bibfnamefont {E.}~\bibnamefont
  {Lim}}, \bibinfo {author} {\bibfnamefont {K.~A.}\ \bibnamefont {Peterson}},
  \bibinfo {author} {\bibfnamefont {G.~M.}\ \bibnamefont {Su}}, and\ \bibinfo
  {author} {\bibfnamefont {M.~L.}\ \bibnamefont {Chabinyc}},\ }\bibfield
  {title} {\bibinfo {title} {Thermoelectric properties of
  poly(3-hexylthiophene) ({P3HT}) doped with
  2,3,5,6-tetrafluoro-7,7,8,8-tetracyanoquinodimethane ({F4TCNQ}) by
  vapor-phase infiltration},\ }\href
  {https://doi.org/10.1021/acs.chemmater.7b04849} {\bibfield  {journal}
  {\bibinfo  {journal} {Chemistry of Materials}\ }\textbf {\bibinfo {volume}
  {30}},\ \bibinfo {pages} {998} (\bibinfo {year} {2018})},\ \Eprint
  {https://arxiv.org/abs/https://doi.org/10.1021/acs.chemmater.7b04849}
  {https://doi.org/10.1021/acs.chemmater.7b04849} \BibitemShut {NoStop}%
\bibitem [{\citenamefont {Kronik}\ and\ \citenamefont
  {Shapira}(1999)}]{KRONIK_SPV}%
  \BibitemOpen
  \bibfield  {author} {\bibinfo {author} {\bibfnamefont {L.}~\bibnamefont
  {Kronik}}and\ \bibinfo {author} {\bibfnamefont {Y.}~\bibnamefont {Shapira}},\
  }\bibfield  {title} {\bibinfo {title} {Surface photovoltage phenomena:
  theory, experiment, and applications},\ }\href
  {https://doi.org/https://doi.org/10.1016/S0167-5729(99)00002-3} {\bibfield
  {journal} {\bibinfo  {journal} {Surface Science Reports}\ }\textbf {\bibinfo
  {volume} {37}},\ \bibinfo {pages} {1} (\bibinfo {year} {1999})}\BibitemShut
  {NoStop}%
\bibitem [{\citenamefont {Li}\ \emph {et~al.}(2020)\citenamefont {Li},
  \citenamefont {Choudhry}, \citenamefont {Ranasinghe}, \citenamefont
  {Ackerman},\ and\ \citenamefont {Liao}}]{li2020probing}%
  \BibitemOpen
  \bibfield  {author} {\bibinfo {author} {\bibfnamefont {Y.}~\bibnamefont
  {Li}}, \bibinfo {author} {\bibfnamefont {U.}~\bibnamefont {Choudhry}},
  \bibinfo {author} {\bibfnamefont {J.}~\bibnamefont {Ranasinghe}}, \bibinfo
  {author} {\bibfnamefont {A.}~\bibnamefont {Ackerman}}, and\ \bibinfo {author}
  {\bibfnamefont {B.}~\bibnamefont {Liao}},\ }\bibfield  {title} {\bibinfo
  {title} {Probing surface photovoltage effect using photoassisted secondary
  electron emission},\ }\href {https://doi.org/10.1021/acs.jpca.0c02543}
  {\bibfield  {journal} {\bibinfo  {journal} {The Journal of Physical Chemistry
  A}\ }\textbf {\bibinfo {volume} {124}},\ \bibinfo {pages} {5246} (\bibinfo
  {year} {2020})},\ \bibinfo {note} {pMID: 32492349},\ \Eprint
  {https://arxiv.org/abs/https://doi.org/10.1021/acs.jpca.0c02543}
  {https://doi.org/10.1021/acs.jpca.0c02543} \BibitemShut {NoStop}%
\bibitem [{\citenamefont {Toer}\ \emph {et~al.}(1998)\citenamefont {Toer},
  \citenamefont {Reimer}, \citenamefont {MacAdam}, \citenamefont {Samuels},
  \citenamefont {Hawkes}, \citenamefont {Schawlow}, \citenamefont {Shimoda},
  \citenamefont {Siegman}, \citenamefont {Tamir},\ and\ \citenamefont
  {Lotasch}}]{Reimer}%
  \BibitemOpen
  \bibfield  {author} {\bibinfo {author} {\bibfnamefont {P.}~\bibnamefont
  {Toer}}, \bibinfo {author} {\bibfnamefont {L.}~\bibnamefont {Reimer}},
  \bibinfo {author} {\bibfnamefont {D.}~\bibnamefont {MacAdam}}, \bibinfo
  {author} {\bibfnamefont {W.}~\bibnamefont {Samuels}}, \bibinfo {author}
  {\bibfnamefont {P.}~\bibnamefont {Hawkes}}, \bibinfo {author} {\bibfnamefont
  {A.}~\bibnamefont {Schawlow}}, \bibinfo {author} {\bibfnamefont
  {K.}~\bibnamefont {Shimoda}}, \bibinfo {author} {\bibfnamefont
  {A.}~\bibnamefont {Siegman}}, \bibinfo {author} {\bibfnamefont
  {T.}~\bibnamefont {Tamir}}, and\ \bibinfo {author} {\bibfnamefont
  {H.}~\bibnamefont {Lotasch}},\ }\href
  {https://books.google.com/books?id=0Fm3T6F6\_LEC} {\emph {\bibinfo {title}
  {Scanning Electron Microscopy: Physics of Image Formation and
  Microanalysis}}},\ Springer Series in Optical Sciences\ (\bibinfo
  {publisher} {Springer},\ \bibinfo {year} {1998})\BibitemShut {NoStop}%
\bibitem [{\citenamefont {Shaw}\ \emph {et~al.}(2008)\citenamefont {Shaw},
  \citenamefont {Ruseckas},\ and\ \citenamefont {Samuel}}]{shaw2008exciton}%
  \BibitemOpen
  \bibfield  {author} {\bibinfo {author} {\bibfnamefont {P.~E.}\ \bibnamefont
  {Shaw}}, \bibinfo {author} {\bibfnamefont {A.}~\bibnamefont {Ruseckas}}, and\
  \bibinfo {author} {\bibfnamefont {I.~D.~W.}\ \bibnamefont {Samuel}},\
  }\bibfield  {title} {\bibinfo {title} {Exciton diffusion measurements in
  poly(3-hexylthiophene)},\ }\href
  {https://doi.org/https://doi.org/10.1002/adma.200800982} {\bibfield
  {journal} {\bibinfo  {journal} {Advanced Materials}\ }\textbf {\bibinfo
  {volume} {20}},\ \bibinfo {pages} {3516} (\bibinfo {year} {2008})},\ \Eprint
  {https://arxiv.org/abs/https://onlinelibrary.wiley.com/doi/pdf/10.1002/adma.200800982}
  {https://onlinelibrary.wiley.com/doi/pdf/10.1002/adma.200800982} \BibitemShut
  {NoStop}%
\bibitem [{\citenamefont {Savenije}\ \emph {et~al.}(2013)\citenamefont
  {Savenije}, \citenamefont {Ferguson}, \citenamefont {Kopidakis},\ and\
  \citenamefont {Rumbles}}]{savenije2013revealing}%
  \BibitemOpen
  \bibfield  {author} {\bibinfo {author} {\bibfnamefont {T.~J.}\ \bibnamefont
  {Savenije}}, \bibinfo {author} {\bibfnamefont {A.~J.}\ \bibnamefont
  {Ferguson}}, \bibinfo {author} {\bibfnamefont {N.}~\bibnamefont {Kopidakis}},
  and\ \bibinfo {author} {\bibfnamefont {G.}~\bibnamefont {Rumbles}},\
  }\bibfield  {title} {\bibinfo {title} {Revealing the dynamics of charge
  carriers in polymer:fullerene blends using photoinduced time-resolved
  microwave conductivity},\ }\href {https://doi.org/10.1021/jp406706u}
  {\bibfield  {journal} {\bibinfo  {journal} {The Journal of Physical Chemistry
  C}\ }\textbf {\bibinfo {volume} {117}},\ \bibinfo {pages} {24085} (\bibinfo
  {year} {2013})},\ \Eprint
  {https://arxiv.org/abs/https://doi.org/10.1021/jp406706u}
  {https://doi.org/10.1021/jp406706u} \BibitemShut {NoStop}%
\bibitem [{\citenamefont {Liao}\ \emph {et~al.}(2016)\citenamefont {Liao},
  \citenamefont {Maznev}, \citenamefont {Nelson},\ and\ \citenamefont
  {Chen}}]{liao2016photo}%
  \BibitemOpen
  \bibfield  {author} {\bibinfo {author} {\bibfnamefont {B.}~\bibnamefont
  {Liao}}, \bibinfo {author} {\bibfnamefont {A.~A.}\ \bibnamefont {Maznev}},
  \bibinfo {author} {\bibfnamefont {K.~A.}\ \bibnamefont {Nelson}}, and\
  \bibinfo {author} {\bibfnamefont {G.}~\bibnamefont {Chen}},\ }\bibfield
  {title} {\bibinfo {title} {Photo-excited charge carriers suppress
  sub-terahertz phonon mode in silicon at room temperature},\ }\href
  {https://doi.org/10.1038/ncomms13174} {\bibfield  {journal} {\bibinfo
  {journal} {Nature Communications}\ }\textbf {\bibinfo {volume} {7}},\
  \bibinfo {pages} {13174} (\bibinfo {year} {2016})}\BibitemShut {NoStop}%
\bibitem [{\citenamefont {Gurau}\ \emph {et~al.}(2007)\citenamefont {Gurau},
  \citenamefont {Delongchamp}, \citenamefont {Vogel}, \citenamefont {Lin},
  \citenamefont {Fischer}, \citenamefont {Sambasivan},\ and\ \citenamefont
  {Richter}}]{gurau2007measuring}%
  \BibitemOpen
  \bibfield  {author} {\bibinfo {author} {\bibfnamefont {M.~C.}\ \bibnamefont
  {Gurau}}, \bibinfo {author} {\bibfnamefont {D.~M.}\ \bibnamefont
  {Delongchamp}}, \bibinfo {author} {\bibfnamefont {B.~M.}\ \bibnamefont
  {Vogel}}, \bibinfo {author} {\bibfnamefont {E.~K.}\ \bibnamefont {Lin}},
  \bibinfo {author} {\bibfnamefont {D.~A.}\ \bibnamefont {Fischer}}, \bibinfo
  {author} {\bibfnamefont {S.}~\bibnamefont {Sambasivan}}, and\ \bibinfo
  {author} {\bibfnamefont {L.~J.}\ \bibnamefont {Richter}},\ }\bibfield
  {title} {\bibinfo {title} {Measuring molecular order in
  poly(3-alkylthiophene) thin films with polarizing spectroscopies},\ }\href
  {https://doi.org/10.1021/la0618972} {\bibfield  {journal} {\bibinfo
  {journal} {Langmuir}\ }\textbf {\bibinfo {volume} {23}},\ \bibinfo {pages}
  {834} (\bibinfo {year} {2007})},\ \bibinfo {note} {pMID: 17209641},\ \Eprint
  {https://arxiv.org/abs/https://doi.org/10.1021/la0618972}
  {https://doi.org/10.1021/la0618972} \BibitemShut {NoStop}%
\bibitem [{\citenamefont {Clark}\ \emph {et~al.}(2007)\citenamefont {Clark},
  \citenamefont {Silva}, \citenamefont {Friend},\ and\ \citenamefont
  {Spano}}]{clark2007role}%
  \BibitemOpen
  \bibfield  {author} {\bibinfo {author} {\bibfnamefont {J.}~\bibnamefont
  {Clark}}, \bibinfo {author} {\bibfnamefont {C.}~\bibnamefont {Silva}},
  \bibinfo {author} {\bibfnamefont {R.~H.}\ \bibnamefont {Friend}}, and\
  \bibinfo {author} {\bibfnamefont {F.~C.}\ \bibnamefont {Spano}},\ }\bibfield
  {title} {\bibinfo {title} {Role of intermolecular coupling in the
  photophysics of disordered organic semiconductors: Aggregate emission in
  regioregular polythiophene},\ }\href
  {https://doi.org/10.1103/PhysRevLett.98.206406} {\bibfield  {journal}
  {\bibinfo  {journal} {Phys. Rev. Lett.}\ }\textbf {\bibinfo {volume} {98}},\
  \bibinfo {pages} {206406} (\bibinfo {year} {2007})}\BibitemShut {NoStop}%
\bibitem [{\citenamefont {Datskos}\ \emph {et~al.}(1998)\citenamefont
  {Datskos}, \citenamefont {Rajic},\ and\ \citenamefont
  {Datskou}}]{datskos1998photoinduced}%
  \BibitemOpen
  \bibfield  {author} {\bibinfo {author} {\bibfnamefont {P.~G.}\ \bibnamefont
  {Datskos}}, \bibinfo {author} {\bibfnamefont {S.}~\bibnamefont {Rajic}}, and\
  \bibinfo {author} {\bibfnamefont {I.}~\bibnamefont {Datskou}},\ }\bibfield
  {title} {\bibinfo {title} {Photoinduced and thermal stress in silicon
  microcantilevers},\ }\href {https://doi.org/10.1063/1.121809} {\bibfield
  {journal} {\bibinfo  {journal} {Applied Physics Letters}\ }\textbf {\bibinfo
  {volume} {73}},\ \bibinfo {pages} {2319} (\bibinfo {year} {1998})},\ \Eprint
  {https://arxiv.org/abs/https://doi.org/10.1063/1.121809}
  {https://doi.org/10.1063/1.121809} \BibitemShut {NoStop}%
\bibitem [{\citenamefont {Zhou}\ \emph {et~al.}(2016)\citenamefont {Zhou},
  \citenamefont {You}, \citenamefont {Wang}, \citenamefont {Ku}, \citenamefont
  {Fan}, \citenamefont {Schmidt}, \citenamefont {Rusydi}, \citenamefont
  {Chang}, \citenamefont {Wang}, \citenamefont {Ren}, \citenamefont {Chen},
  \citenamefont {Yuan}, \citenamefont {Chen},\ and\ \citenamefont
  {Wang}}]{zhou2016giant}%
  \BibitemOpen
  \bibfield  {author} {\bibinfo {author} {\bibfnamefont {Y.}~\bibnamefont
  {Zhou}}, \bibinfo {author} {\bibfnamefont {L.}~\bibnamefont {You}}, \bibinfo
  {author} {\bibfnamefont {S.}~\bibnamefont {Wang}}, \bibinfo {author}
  {\bibfnamefont {Z.}~\bibnamefont {Ku}}, \bibinfo {author} {\bibfnamefont
  {H.}~\bibnamefont {Fan}}, \bibinfo {author} {\bibfnamefont {D.}~\bibnamefont
  {Schmidt}}, \bibinfo {author} {\bibfnamefont {A.}~\bibnamefont {Rusydi}},
  \bibinfo {author} {\bibfnamefont {L.}~\bibnamefont {Chang}}, \bibinfo
  {author} {\bibfnamefont {L.}~\bibnamefont {Wang}}, \bibinfo {author}
  {\bibfnamefont {P.}~\bibnamefont {Ren}}, \bibinfo {author} {\bibfnamefont
  {L.}~\bibnamefont {Chen}}, \bibinfo {author} {\bibfnamefont {G.}~\bibnamefont
  {Yuan}}, \bibinfo {author} {\bibfnamefont {L.}~\bibnamefont {Chen}}, and\
  \bibinfo {author} {\bibfnamefont {J.}~\bibnamefont {Wang}},\ }\bibfield
  {title} {\bibinfo {title} {Giant photostriction in organic--inorganic lead
  halide perovskites},\ }\href {https://doi.org/10.1038/ncomms11193} {\bibfield
   {journal} {\bibinfo  {journal} {Nature Communications}\ }\textbf {\bibinfo
  {volume} {7}},\ \bibinfo {pages} {11193} (\bibinfo {year}
  {2016})}\BibitemShut {NoStop}%
\bibitem [{\citenamefont {Chen\:}\ and\ \citenamefont
  {Yi}(2021)}]{chen2021photostrictive}%
  \BibitemOpen
  \bibfield  {author} {\bibinfo {author} {\bibfnamefont {C.}~\bibnamefont
  {Chen\:}}and\ \bibinfo {author} {\bibfnamefont {Z.}~\bibnamefont {Yi}},\
  }\bibfield  {title} {\bibinfo {title} {Photostrictive effect:
  Characterization techniques, materials, and applications},\ }\href
  {https://doi.org/https://doi.org/10.1002/adfm.202010706} {\bibfield
  {journal} {\bibinfo  {journal} {Advanced Functional Materials}\ }\textbf
  {\bibinfo {volume} {31}},\ \bibinfo {pages} {2010706} (\bibinfo {year}
  {2021})},\ \Eprint
  {https://arxiv.org/abs/https://onlinelibrary.wiley.com/doi/pdf/10.1002/adfm.202010706}
  {https://onlinelibrary.wiley.com/doi/pdf/10.1002/adfm.202010706} \BibitemShut
  {NoStop}%
\bibitem [{\citenamefont {Street}(2008)}]{street2008bias}%
  \BibitemOpen
  \bibfield  {author} {\bibinfo {author} {\bibfnamefont {R.~A.}\ \bibnamefont
  {Street}},\ }\bibfield  {title} {\bibinfo {title} {Bias-induced change in
  effective mobility observed in polymer transistors},\ }\href
  {https://doi.org/10.1103/PhysRevB.77.165311} {\bibfield  {journal} {\bibinfo
  {journal} {Phys. Rev. B}\ }\textbf {\bibinfo {volume} {77}},\ \bibinfo
  {pages} {165311} (\bibinfo {year} {2008})}\BibitemShut {NoStop}%
\bibitem [{\citenamefont {Kaake}\ \emph {et~al.}(2012)\citenamefont {Kaake},
  \citenamefont {Welch}, \citenamefont {Moses}, \citenamefont {Bazan},\ and\
  \citenamefont {Heeger}}]{kaake2012influence}%
  \BibitemOpen
  \bibfield  {author} {\bibinfo {author} {\bibfnamefont {L.~G.}\ \bibnamefont
  {Kaake}}, \bibinfo {author} {\bibfnamefont {G.~C.}\ \bibnamefont {Welch}},
  \bibinfo {author} {\bibfnamefont {D.}~\bibnamefont {Moses}}, \bibinfo
  {author} {\bibfnamefont {G.~C.}\ \bibnamefont {Bazan}}, and\ \bibinfo
  {author} {\bibfnamefont {A.~J.}\ \bibnamefont {Heeger}},\ }\bibfield  {title}
  {\bibinfo {title} {Influence of processing additives on charge-transfer time
  scales and sound velocity in organic bulk heterojunction films},\ }\href
  {https://doi.org/10.1021/jz300365b} {\bibfield  {journal} {\bibinfo
  {journal} {The Journal of Physical Chemistry Letters}\ }\textbf {\bibinfo
  {volume} {3}},\ \bibinfo {pages} {1253} (\bibinfo {year} {2012})},\ \bibinfo
  {note} {pMID: 26286767},\ \Eprint
  {https://arxiv.org/abs/https://doi.org/10.1021/jz300365b}
  {https://doi.org/10.1021/jz300365b} \BibitemShut {NoStop}%
\bibitem [{\citenamefont {Kanner}\ \emph {et~al.}(1990)\citenamefont {Kanner},
  \citenamefont {Vardeny},\ and\ \citenamefont {Hess}}]{kanner1990picosecond}%
  \BibitemOpen
  \bibfield  {author} {\bibinfo {author} {\bibfnamefont {G.~S.}\ \bibnamefont
  {Kanner}}, \bibinfo {author} {\bibfnamefont {Z.~V.}\ \bibnamefont {Vardeny}},
  and\ \bibinfo {author} {\bibfnamefont {B.~C.}\ \bibnamefont {Hess}},\
  }\bibfield  {title} {\bibinfo {title} {Picosecond acoustics in polythiophene
  thin films},\ }\href {https://doi.org/10.1103/PhysRevB.42.5403} {\bibfield
  {journal} {\bibinfo  {journal} {Phys. Rev. B}\ }\textbf {\bibinfo {volume}
  {42}},\ \bibinfo {pages} {5403} (\bibinfo {year} {1990})}\BibitemShut
  {NoStop}%
\end{thebibliography}%


\begin{thebibliography}{8}%
\makeatletter
\providecommand \@ifxundefined [1]{%
 \@ifx{#1\undefined}
}%
\providecommand \@ifnum [1]{%
 \ifnum #1\expandafter \@firstoftwo
 \else \expandafter \@secondoftwo
 \fi
}%
\providecommand \@ifx [1]{%
 \ifx #1\expandafter \@firstoftwo
 \else \expandafter \@secondoftwo
 \fi
}%
\providecommand \natexlab [1]{#1}%
\providecommand \enquote  [1]{``#1''}%
\providecommand \bibnamefont  [1]{#1}%
\providecommand \bibfnamefont [1]{#1}%
\providecommand \citenamefont [1]{#1}%
\providecommand \href@noop [0]{\@secondoftwo}%
\providecommand \href [0]{\begingroup \@sanitize@url \@href}%
\providecommand \@href[1]{\@@startlink{#1}\@@href}%
\providecommand \@@href[1]{\endgroup#1\@@endlink}%
\providecommand \@sanitize@url [0]{\catcode `\\12\catcode `\$12\catcode
  `\&12\catcode `\#12\catcode `\^12\catcode `\_12\catcode `\%12\relax}%
\providecommand \@@startlink[1]{}%
\providecommand \@@endlink[0]{}%
\providecommand \url  [0]{\begingroup\@sanitize@url \@url }%
\providecommand \@url [1]{\endgroup\@href {#1}{\urlprefix }}%
\providecommand \urlprefix  [0]{URL }%
\providecommand \Eprint [0]{\href }%
\providecommand \doibase [0]{https://doi.org/}%
\providecommand \selectlanguage [0]{\@gobble}%
\providecommand \bibinfo  [0]{\@secondoftwo}%
\providecommand \bibfield  [0]{\@secondoftwo}%
\providecommand \translation [1]{[#1]}%
\providecommand \BibitemOpen [0]{}%
\providecommand \bibitemStop [0]{}%
\providecommand \bibitemNoStop [0]{.\EOS\space}%
\providecommand \EOS [0]{\spacefactor3000\relax}%
\providecommand \BibitemShut  [1]{\csname bibitem#1\endcsname}%
\let\auto@bib@innerbib\@empty
\bibitem [{\citenamefont {Howard}\ \emph {et~al.}(2010)\citenamefont {Howard},
  \citenamefont {Mauer}, \citenamefont {Meister},\ and\ \citenamefont
  {Laquai}}]{howard2010effect}%
  \BibitemOpen
  \bibfield  {author} {\bibinfo {author} {\bibfnamefont {I.~A.}\ \bibnamefont
  {Howard}}, \bibinfo {author} {\bibfnamefont {R.}~\bibnamefont {Mauer}},
  \bibinfo {author} {\bibfnamefont {M.}~\bibnamefont {Meister}}, and\ \bibinfo
  {author} {\bibfnamefont {F.}~\bibnamefont {Laquai}},\ }\bibfield  {title}
  {\bibinfo {title} {Effect of morphology on ultrafast free carrier generation
  in polythiophene:fullerene organic solar cells},\ }\href
  {https://doi.org/10.1021/ja105260d} {\bibfield  {journal} {\bibinfo
  {journal} {Journal of the American Chemical Society}\ }\textbf {\bibinfo
  {volume} {132}},\ \bibinfo {pages} {14866} (\bibinfo {year} {2010})},\
  \bibinfo {note} {pMID: 20923187},\ \Eprint
  {https://arxiv.org/abs/https://doi.org/10.1021/ja105260d}
  {https://doi.org/10.1021/ja105260d} \BibitemShut {NoStop}%
\bibitem [{\citenamefont {Clark}\ \emph {et~al.}(2009)\citenamefont {Clark},
  \citenamefont {Chang}, \citenamefont {Spano}, \citenamefont {Friend},\ and\
  \citenamefont {Silva}}]{clark2009determining}%
  \BibitemOpen
  \bibfield  {author} {\bibinfo {author} {\bibfnamefont {J.}~\bibnamefont
  {Clark}}, \bibinfo {author} {\bibfnamefont {J.-F.}\ \bibnamefont {Chang}},
  \bibinfo {author} {\bibfnamefont {F.~C.}\ \bibnamefont {Spano}}, \bibinfo
  {author} {\bibfnamefont {R.~H.}\ \bibnamefont {Friend}}, and\ \bibinfo
  {author} {\bibfnamefont {C.}~\bibnamefont {Silva}},\ }\bibfield  {title}
  {\bibinfo {title} {Determining exciton bandwidth and film microstructure in
  polythiophene films using linear absorption spectroscopy},\ }\href
  {https://doi.org/10.1063/1.3110904} {\bibfield  {journal} {\bibinfo
  {journal} {Applied Physics Letters}\ }\textbf {\bibinfo {volume} {94}},\
  \bibinfo {pages} {163306} (\bibinfo {year} {2009})},\ \Eprint
  {https://arxiv.org/abs/https://doi.org/10.1063/1.3110904}
  {https://doi.org/10.1063/1.3110904} \BibitemShut {NoStop}%
\bibitem [{\citenamefont {Kim}\ \emph {et~al.}(2021)\citenamefont {Kim},
  \citenamefont {Moon},\ and\ \citenamefont {Minnich}}]{Kim_aSi}%
  \BibitemOpen
  \bibfield  {author} {\bibinfo {author} {\bibfnamefont {T.}~\bibnamefont
  {Kim}}, \bibinfo {author} {\bibfnamefont {J.}~\bibnamefont {Moon}}, and\
  \bibinfo {author} {\bibfnamefont {A.~J.}\ \bibnamefont {Minnich}},\
  }\bibfield  {title} {\bibinfo {title} {Origin of micrometer-scale propagation
  lengths of heat-carrying acoustic excitations in amorphous silicon},\ }\href
  {https://doi.org/10.1103/PhysRevMaterials.5.065602} {\bibfield  {journal}
  {\bibinfo  {journal} {Phys. Rev. Materials}\ }\textbf {\bibinfo {volume}
  {5}},\ \bibinfo {pages} {065602} (\bibinfo {year} {2021})}\BibitemShut
  {NoStop}%
\bibitem [{\citenamefont {Duda}\ \emph {et~al.}(2013)\citenamefont {Duda},
  \citenamefont {Hopkins}, \citenamefont {Shen},\ and\ \citenamefont
  {Gupta}}]{duda2013thermal}%
  \BibitemOpen
  \bibfield  {author} {\bibinfo {author} {\bibfnamefont {J.~C.}\ \bibnamefont
  {Duda}}, \bibinfo {author} {\bibfnamefont {P.~E.}\ \bibnamefont {Hopkins}},
  \bibinfo {author} {\bibfnamefont {Y.}~\bibnamefont {Shen}}, and\ \bibinfo
  {author} {\bibfnamefont {M.~C.}\ \bibnamefont {Gupta}},\ }\bibfield  {title}
  {\bibinfo {title} {Thermal transport in organic semiconducting polymers},\
  }\href {https://doi.org/10.1063/1.4812234} {\bibfield  {journal} {\bibinfo
  {journal} {Applied Physics Letters}\ }\textbf {\bibinfo {volume} {102}},\
  \bibinfo {pages} {251912} (\bibinfo {year} {2013})},\ \Eprint
  {https://arxiv.org/abs/https://doi.org/10.1063/1.4812234}
  {https://doi.org/10.1063/1.4812234} \BibitemShut {NoStop}%
\bibitem [{\citenamefont {Patr{\'{\i}}cio}\ \emph {et~al.}(2006)\citenamefont
  {Patr{\'{\i}}cio}, \citenamefont {Calado}, \citenamefont {de~Oliveira},
  \citenamefont {Righi}, \citenamefont {Neves}, \citenamefont {Silva},\ and\
  \citenamefont {Cury}}]{Patricio_Cp_2006}%
  \BibitemOpen
  \bibfield  {author} {\bibinfo {author} {\bibfnamefont {P.~S.~O.}\
  \bibnamefont {Patr{\'{\i}}cio}}, \bibinfo {author} {\bibfnamefont {H.~D.~R.}\
  \bibnamefont {Calado}}, \bibinfo {author} {\bibfnamefont {F.~A.~C.}\
  \bibnamefont {de~Oliveira}}, \bibinfo {author} {\bibfnamefont
  {A.}~\bibnamefont {Righi}}, \bibinfo {author} {\bibfnamefont {B.~R.~A.}\
  \bibnamefont {Neves}}, \bibinfo {author} {\bibfnamefont {G.~G.}\ \bibnamefont
  {Silva}}, and\ \bibinfo {author} {\bibfnamefont {L.~A.}\ \bibnamefont
  {Cury}},\ }\bibfield  {title} {\bibinfo {title} {Correlation between thermal,
  optical and morphological properties of heterogeneous blends of
  poly(3-hexylthiophene) and thermoplastic polyurethane},\ }\href
  {https://doi.org/10.1088/0953-8984/18/32/002} {\bibfield  {journal} {\bibinfo
   {journal} {Journal of Physics: Condensed Matter}\ }\textbf {\bibinfo
  {volume} {18}},\ \bibinfo {pages} {7529} (\bibinfo {year}
  {2006})}\BibitemShut {NoStop}%
\bibitem [{\citenamefont {Sun}\ \emph {et~al.}(2013)\citenamefont {Sun},
  \citenamefont {Han},\ and\ \citenamefont {Liu}}]{Sun_ChineseBulletin}%
  \BibitemOpen
  \bibfield  {author} {\bibinfo {author} {\bibfnamefont {Y.}~\bibnamefont
  {Sun}}, \bibinfo {author} {\bibfnamefont {Y.}~\bibnamefont {Han}}, and\
  \bibinfo {author} {\bibfnamefont {J.}~\bibnamefont {Liu}},\ }\bibfield
  {title} {\bibinfo {title} {Controlling pcbm aggregation in p3ht/pcbm film by
  a selective solvent vapor annealing},\ }\href
  {https://doi.org/10.1007/s11434-013-5944-6} {\bibfield  {journal} {\bibinfo
  {journal} {Chinese Science Bulletin}\ }\textbf {\bibinfo {volume} {58}},\
  \bibinfo {pages} {2767} (\bibinfo {year} {2013})}\BibitemShut {NoStop}%
\bibitem [{\citenamefont {Root}\ \emph {et~al.}(2017)\citenamefont {Root},
  \citenamefont {Savagatrup}, \citenamefont {Printz}, \citenamefont
  {Rodriquez},\ and\ \citenamefont {Lipomi}}]{root2017mechanical}%
  \BibitemOpen
  \bibfield  {author} {\bibinfo {author} {\bibfnamefont {S.~E.}\ \bibnamefont
  {Root}}, \bibinfo {author} {\bibfnamefont {S.}~\bibnamefont {Savagatrup}},
  \bibinfo {author} {\bibfnamefont {A.~D.}\ \bibnamefont {Printz}}, \bibinfo
  {author} {\bibfnamefont {D.}~\bibnamefont {Rodriquez}}, and\ \bibinfo
  {author} {\bibfnamefont {D.~J.}\ \bibnamefont {Lipomi}},\ }\bibfield  {title}
  {\bibinfo {title} {Mechanical properties of organic semiconductors for
  stretchable, highly flexible, and mechanically robust electronics},\ }\href
  {https://doi.org/10.1021/acs.chemrev.7b00003} {\bibfield  {journal} {\bibinfo
   {journal} {Chemical Reviews}\ }\textbf {\bibinfo {volume} {117}},\ \bibinfo
  {pages} {6467} (\bibinfo {year} {2017})},\ \bibinfo {note} {pMID: 28343389},\
  \Eprint {https://arxiv.org/abs/https://doi.org/10.1021/acs.chemrev.7b00003}
  {https://doi.org/10.1021/acs.chemrev.7b00003} \BibitemShut {NoStop}%
\bibitem [{\citenamefont {Jaglarz}\ \emph {et~al.}(2016)\citenamefont
  {Jaglarz}, \citenamefont {Nosidlak},\ and\ \citenamefont
  {Wolska}}]{jaglarz2016thermo}%
  \BibitemOpen
  \bibfield  {author} {\bibinfo {author} {\bibfnamefont {J.}~\bibnamefont
  {Jaglarz}}, \bibinfo {author} {\bibfnamefont {N.}~\bibnamefont {Nosidlak}},
  and\ \bibinfo {author} {\bibfnamefont {N.}~\bibnamefont {Wolska}},\
  }\bibfield  {title} {\bibinfo {title} {Thermo-optical properties of conducted
  polythiophene polymer films used in electroluminescent devices},\ }\href
  {https://doi.org/10.1007/s11082-016-0649-0} {\bibfield  {journal} {\bibinfo
  {journal} {Optical and Quantum Electronics}\ }\textbf {\bibinfo {volume}
  {48}},\ \bibinfo {pages} {392} (\bibinfo {year} {2016})}\BibitemShut
  {NoStop}%
\end{thebibliography}%

\end{document}


\title{Supplementary Information: Transient strain induced electronic structure modulation in a semiconducting polymer imaged by scanning ultrafast electron microscopy}

\author{Taeyong Kim}
\affiliation{Department of Mechanical Engineering, University of California, Santa Barbara, CA 93106, USA}

\author{Saejin Oh}
\affiliation{Department of Chemistry and Biochemistry, University of California, Santa Barbara, CA 93106, USA}

\author{Usama Choudhry}
\affiliation{Department of Mechanical Engineering, University of California, Santa Barbara, CA 93106, USA}

\author{Carl Meinhart}
\affiliation{Department of Mechanical Engineering, University of California, Santa Barbara, CA 93106, USA}

\author{Michael L. Chabinyc}
\email{chabiny@ucsb.edu }
\affiliation{Materials Department, University of California, Santa Barbara, CA 93106, USA}

\author{Bolin Liao}
\email{bliao@ucsb.edu} \affiliation{Department of Mechanical Engineering, University of California, Santa Barbara, CA 93106, USA}


\date{\today}

{    \global\let\newpagegood\newpage
    \global\let\newpage\relax
\maketitle}
\clearpage
\section{Additional SUEM Contrast Images of Neat P3HT with various optical pump powers}
Additional representative SUEM contrast images of neat P3HT taken at different optical excitation powers are presented here. At all the optical powers used in this work, a similar ring-shaped profile which includes dark and bright secondary electron contrast is clearly visible, as shown in both data sets below.

\begin{figure}[hbt!]
\centering
\includegraphics[width=\columnwidth,keepaspectratio]{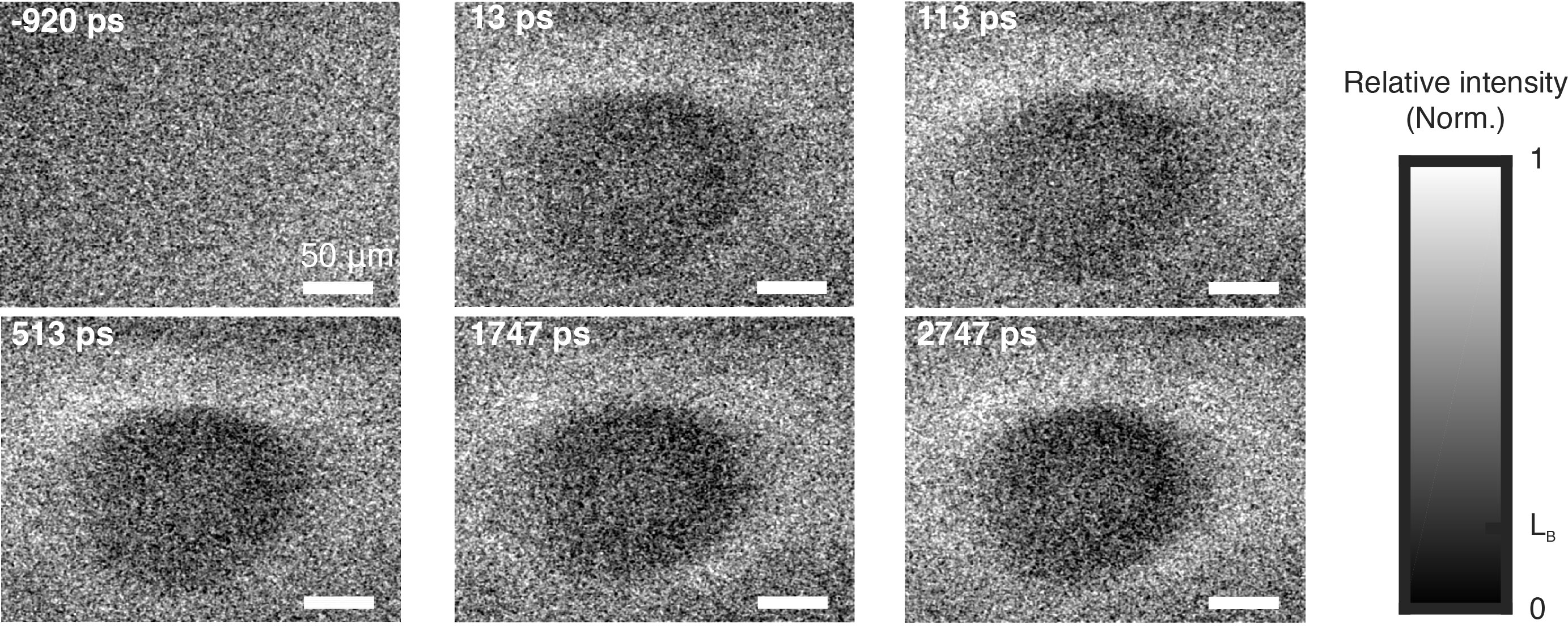}
\caption{\textbf{Additional SUEM Contrast Images of Neat P3HT with an Optical Pump Power of 700 $\mu$W.} The procedures for data acquisition and image processing are identical to those described in the main text, except the used primary electron current of $\sim 18$ pA. The color bar indicates the normalized intensity distribution at each time delay, in which $L_B$ labels the average background intensity level.}
\label{sfig:diffimgP3HT700uW.jpg}
\end{figure}

\begin{figure}[hbt!]
\centering
\includegraphics[width=\columnwidth,keepaspectratio]{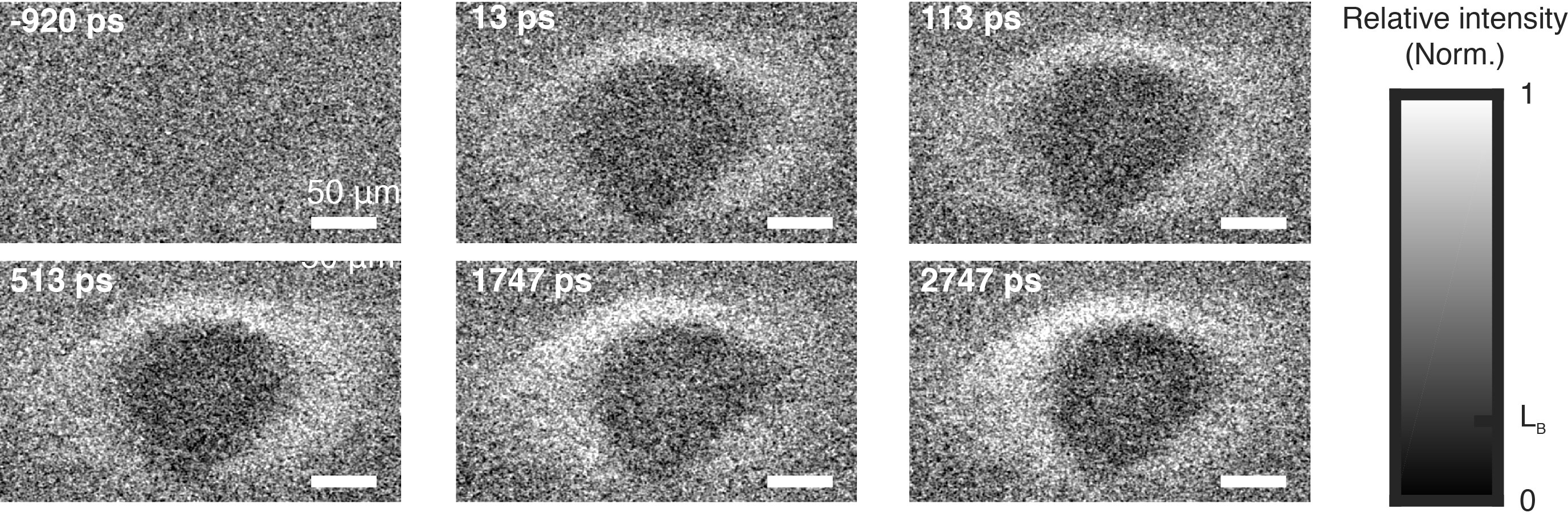}
\caption{\textbf{Additional SUEM Contrast Images of Neat P3HT with an Optical Pump Power of 900 $\mu$W.}  The procedures for data acquisition and image processing are identical to those described in the main text, except the used primary electron current of $\sim 18$ pA. The color bar indicates the normalized intensity distribution at each time delay, in which $L_B$ labels the average background intensity level.}
\label{sfig:diffimgP3HT900uW.jpg}
\end{figure}

\clearpage

\section{SUEM Contrast Images of P3HT:PCBM Blend}

\begin{figure}[hbt!]
\centering
\includegraphics[width=\columnwidth,keepaspectratio]{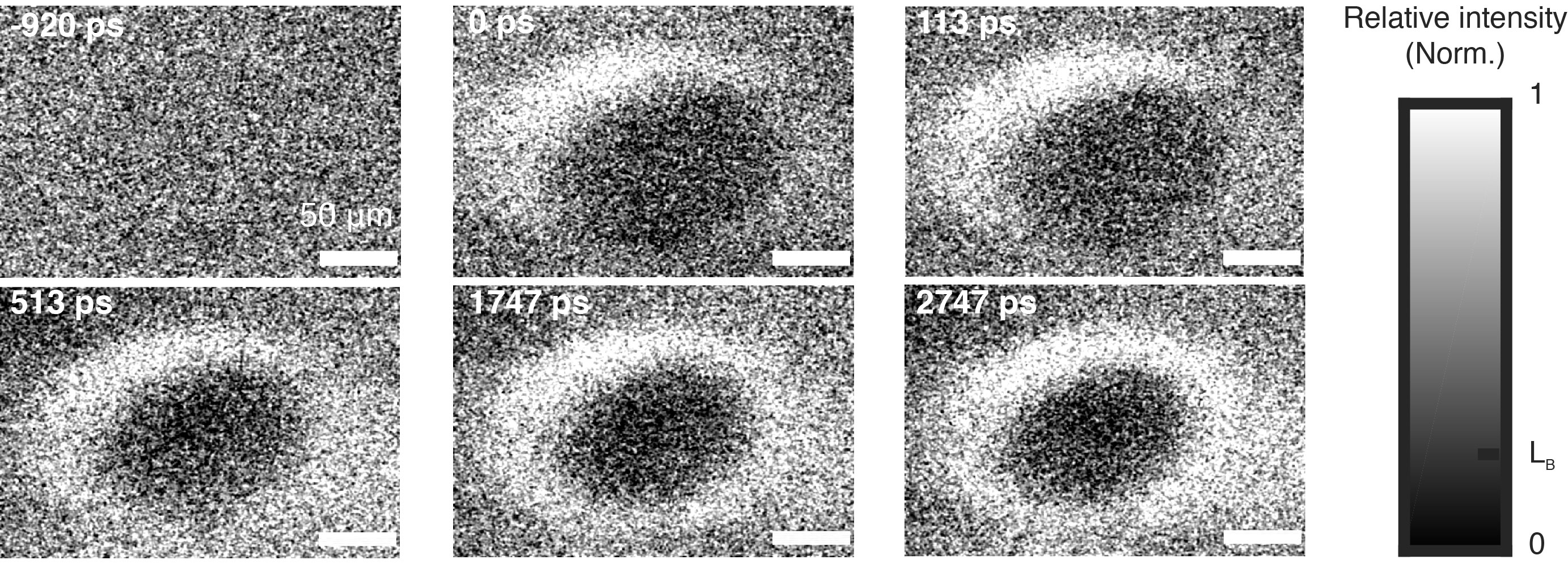}
\caption{\textbf{SUEM Contrast Images of P3HT:PCBM Blend with an Optical Pump Power of 800 $\mu$W.} The procedures for data acquisition and image processing are identical to those described in the main text. the primary electron current of $\sim 80$ pA was used. The color bar indicates the normalized intensity distribution at each time delay, in which $L_B$ labels the average background intensity level.}
\label{sfig:diffimgPCBM800uW}
\end{figure}

We conducted SUEM measurements of a P3HT/1 wt \% PCBM blend sample and the contrast images are shown in Fig.~\ref{sfig:diffimgPCBM800uW}. The optical pump power was 800 $\mu$W. Qualitatively, the image contrast between the bright and the dark areas in the blend sample is stronger than that in neat P3HT, indicating that photo-induced stress in the blend is stronger than that in the neat P3HT. The main difference between P3HT and the P3HT/PCBM blend is the type of major photogenerated species in the measurement time window~\cite{howard2010effect}: while a significant amount of excitons remain in the neat P3HT within 2 ns after the photo-exictation, the excitons are strongly suppressed by PCBM and the major photocarriers in the blend are free charges. The observed difference in the SUEM contrast between the two samples signals a stronger electron-lattice coupling associated with the free charges than the excitons in conducting polymers.

\clearpage

\section{UV-VIS spectral analysis}

\begin{figure}[hbt!]
\centering
\includegraphics[width=.6\columnwidth,keepaspectratio]{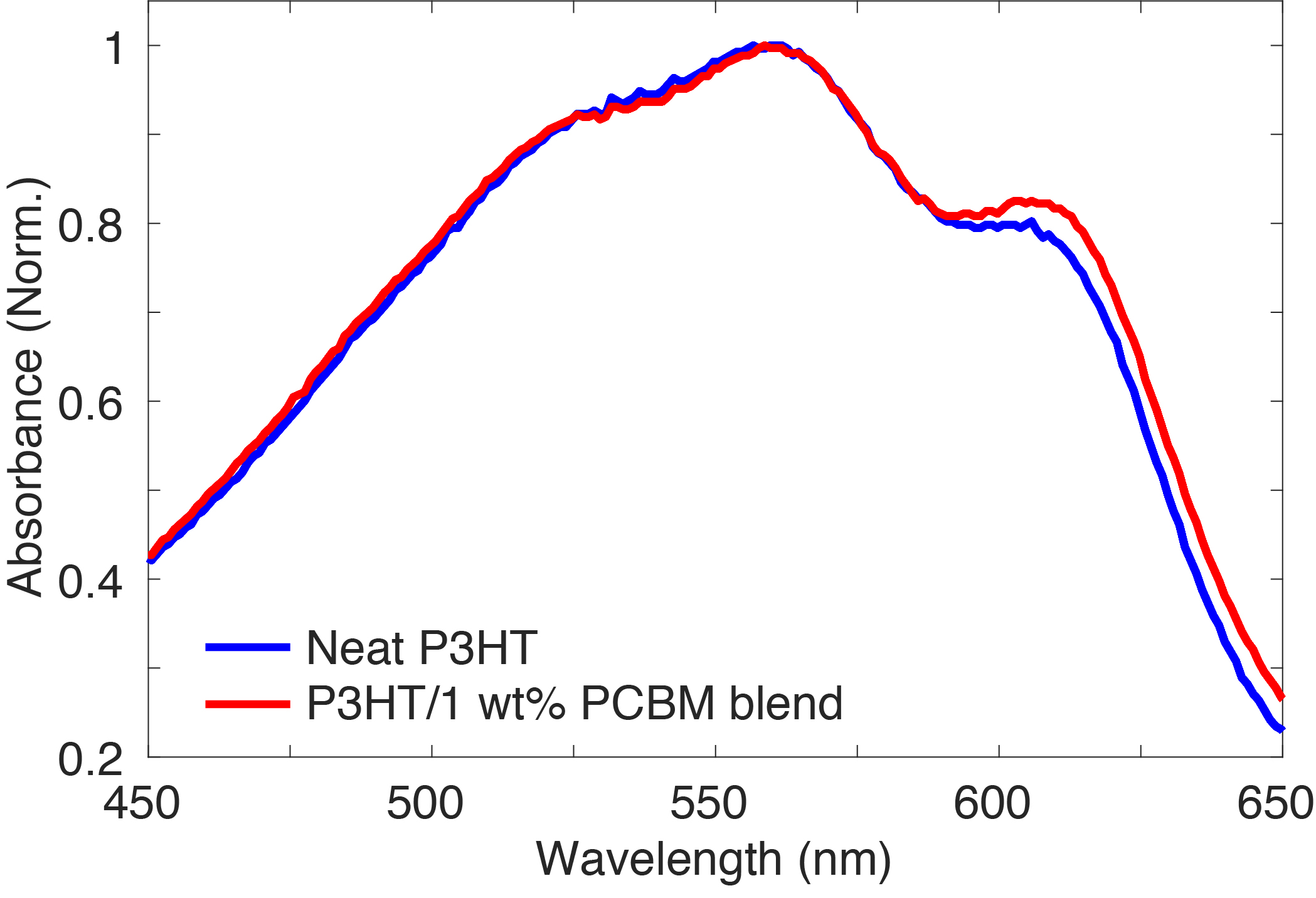}
\caption{\textbf{The UV-VIS absorption spectra for P3HT and P3HT:PCBM thin films} The absorbance was normalized for the two spectra.
The absorption peaks at 620 nm and 560 nm  can be assigned to ordered regions of the film with H-like aggregate character~\cite{clark2009determining}.}  
\label{sfig:UVVIS}
\end{figure}

We measured the ultraviolet-visible (UV-VIS) spectra of the samples to characterize the structural order and optical properties of the samples. The corresponding absorbance versus the wavelength for the P3HT and the P3HT/PCBM blend is given in Figure~\ref{sfig:UVVIS}. In the spectra in Fig.~\ref{sfig:UVVIS}, three peaks are visible at $\sim 520-620$ nm in both of the films and show that there is H-like aggregate character in the ordered regions of the films \cite{clark2009determining}.  There are no substantial changes in the optical spectrum upon addition of a dilute amount of PCBM.

\clearpage

\section{Estimation of the laser induced transient heating}
Laser induced heating of the polymer film can lead to a thermal stress that initiates a dynamic elastic response. The magnitude of the thermal stress can be estimated by considering the absorbed optical energy per pulse and the volume over which the optical energy is deposited. Following Ref.~\cite{Kim_aSi}, the transient temperature rise ($\Delta T$) induced by the absorbed optical energy can be estimated from:

\begin{eqnarray}
\Delta T = \frac{E_{abs}}{C_v V}, \label{eq:Trans_HeatCond}
\end{eqnarray}\\
where $E_{abs}$ is the absorbed pump energy, $V$ is the volume of the sample that absorbs the pump energy, and $C_{v}$ is the volumetric heat capacity. The volume $V = \pi r^2 d$, where $d \approx 100$ nm is the optical penetration depth \cite{duda2013thermal}, and $r$ is the 1/$e^2$ pump beam radius ($\sim 50$ $\mu $m). The volumetric heat capacity was calculated to be 1.43 $J$~$cm^{-3}K^{-1}$ using the literature values of the isobaric specific heat capacity of $1.3$ $J$~$g^{-1}K^{-1}$ \cite{Patricio_Cp_2006} along with the density of $1.1 $ $g~cm^{-3}$ \cite{Sun_ChineseBulletin}. The incident optical pump energy on the sample was calculated using $E_{inc} = \tau P_{pu}/f$ where $\tau \sim 92$\% is the optical transmission of the UV fused-silica window, $P_{pu}$ ($\sim 800$ $\mu$W) is the power measured before the window, and $f$ is the laser repetition rate (1 MHz). The $E_{inc} > E_{abs}$ was estimated to be $\sim 0.7$ nJ. Then the resulting transient temperature rise due to the absorption of a single optical pump pulse is estimated to be $\lesssim 1$ K.

Given the mechanical properties of P3HT~\cite{root2017mechanical} and its linear thermal expansion coefficient of $8.9 \times 10^{-4}$~$K^{-1}$ \cite{jaglarz2016thermo}, our FEM simulation predicted that the radial strain induced by the thermal stress is on the order of $10^{-8}$. This small strain indicates an unphysical radial displacement that is unlikely responsible for the SUEM response we observed. Thus, we believe the photostrictive stress generated as a result of the direct coupling between photocarriers and the polymer chains is a more probable cause for the observed strain profile.

\bibliography{referencesSI}